\documentclass[aps,showpacs,preprintnumbers,amsmath,amssymb]{revtex4}

\usepackage{graphicx}
\usepackage{epstopdf}
\usepackage{latexsym}
\usepackage{amssymb}
\usepackage{amsmath}
\usepackage{color}
\usepackage{float}
\usepackage{mathrsfs}
\usepackage[center]{subfigure}
\usepackage{dcolumn}
\usepackage[dvips]{epsfig}

\usepackage{graphicx}
\usepackage{epstopdf}
\usepackage{latexsym}
\usepackage{amssymb}
\usepackage{amsmath}
\usepackage{color}
\usepackage{mathrsfs}
\usepackage[center]{subfigure}
\usepackage{ifpdf}
\usepackage{epstopdf}
\usepackage{multirow}

\begin{document}
\baselineskip=0.8 cm
\title{{\bf A new asymptotical flat and spherically symmetric solution in the generalized Einstein-Cartan-Kibble-Sciama gravity and gravitational lensing}}

\author{Songbai Chen$^{1,2,3,4}$\footnote{Corresponding author: csb3752@hunnu.edu.cn}, Lu Zhang$^{1,2}$, Jiliang
Jing$^{1,2,3,4}$ \footnote{jljing@hunnu.edu.cn}}

\affiliation{$^{\textit{1}}$Institute of Physics and Department of Physics, Hunan
Normal University,  Changsha, Hunan 410081, People's Republic of
China \\ $^{\textit{2}}$Key Laboratory of Low Dimensional Quantum Structures \\
and Quantum Control of Ministry of Education, Hunan Normal
University, Changsha, Hunan 410081, People's Republic of China\\
$^{\textit{3}}$Synergetic Innovation Center for Quantum Effects and Applications,
Hunan Normal University, Changsha, Hunan 410081, People's Republic
of China\\$ ^4$Center for Gravitation and Cosmology, College of Physical Science and Technology,
Yangzhou University, Yangzhou 225009, China}

\vspace*{0.2cm}
\begin{abstract}
\baselineskip=0.6 cm
\begin{center}
{\bf Abstract}
\end{center}
We firstly present a new asymptotical flat and spherically symmetric solution in the generalized Einstein-Cartan-Kibble-Sciama (ECKS) theory of gravity and then investigate the propagation of photon in this background.  This solution possesses three independent parameters which affect sharply photon sphere, deflection angle of light ray and gravitational lensing. Since the condition of existence of horizons is not inconsistent with that of photon sphere, there exists a special case where there is horizon but no photon sphere in this spacetime. Especially, we find that in this special case, the deflection angle of a light ray near the event horizon tends to a finite value rather than
diverges, which is not explored in other spacetimes. We also study the strong gravitational lensing in this spacetime with the photon sphere and then probe how the spacetime parameters affect the coefficients in the strong field limit.
\end{abstract}

 \maketitle
\newpage
\vspace*{0.2cm}
\section{Introduction}

General relativity is the most beautiful theory of gravity at present and it is a fundamental theoretical setting for the modern astrophysics and cosmology.
However, the observed accelerating expansion of the current Universe \cite{t5,t501,t51,t6,t61} implies that some important ingredients could be missing in this theory. One of ingredients which is absent in Einstein theory is torsion, which is the antisymmetric part of the general affine connection. In the gravity theories with torsion, the gravitational field is described by both of spacetime metric and torsion field, which means that the emergence of torsion will modify the feature of the gravitational interaction.

One of natural extensions of Einstein's theory of gravity is the so-called ECKS theory of gravity \cite{Cartan,Kibble}. In this theory, the curvature and the torsion,  respectively,  are assumed to couple with the energy and momentum and the intrinsic angular momentum of matter.  The gravitational repulsion effect arising from such a spinor-torsion coupling can avoid the formation of spacetime singularities in the region with extremely large densities, for example, in the interior of black holes and the very early stage of Universe \cite{avoid,avoid1,avoid2,avoid3}. In the low densities region, the ECSK theory and Einstein's general relativity give indistinguishable predictions since the contribution from torsion to the Einstein equations is negligibly small. However, the torsion field is not dynamical in the ECSK theory since the torsion equation is an algebraic constraint rather than a partial differential equation, which means that the torsion field outside of matter distribution vanishes because it can not propagate as a wave in the spacetime.

In order to construct a dynamical torsion field, one can generalize ECSK theory by introducing
higher order corrections in Lagrangian \cite{gK1,gK2,gK2q,gK11}.
These coupling terms between the spacetime torsion and curvature yield that both equations of motion for spacetime torsion and curvature are dynamical equations, which ensures that the spacetime torsion can propagate in the spacetime even in the absence of spin of matter. In the frame of Poincar\'{e} gauge theory, all possible quadratic invariants in Lagrangian were constructed for dynamical torsion field \cite{gK1,gK11}, and then the corresponding Reissner-Nordstr\"{o}m type and Reissner-Nordstr\"{o}m de Sitter type solutions have been found in this generalized ECSK theory with quadratic invariants \cite{gK12}.
Motivated by that at the one-loop level the infinity structure of the functional integral in quantised gravitational theory should contain the scalar invariants including fourth order derivatives of the metric,
S. Christensen built all possible fourth order scalar invariants on curved manifolds by using the torsion tensor, Riemann tensor and their derivatives  \cite{gK2}. E. Sezgin \textit{et al} investigated such a kind of higher-derivative theory of gravity with propagating torsion in which there is no ghosts or tachyons \cite{gK2q}. Recently, H. Shabani \textit{et al} \cite{sbh1} modified the action in the usual ECSK theory by adding the  interaction term $\tilde{R}\mathcal{T}$ between curvature scalar $\tilde{R}$ and torsion scalar $\mathcal{T}$. The main reasons of  selecting such a coupling term is as follows: (i) Both of curvature scalar $\tilde{R}$ and torsion scalar $\mathcal{T}$ are important quantities in the theory of gravity with torsion. Especially,  $\mathcal{T}$ plays an import role in New General Relativity (NGR) which is one of the generalizations in teleparallelism with the equivalence outcome as general relativity \cite{TNG1} and then some theoretical models with $\mathcal{T}$ have been also investigated extensively in cosmology to explain the accelerating expansion of Universe. (ii) The direct interaction between the torsion scalar $\mathcal{T}$ and curvature scalar $\tilde{R}$ could exist in the regimes of extreme gravity which are present in a black hole spacetime. (iii) The coupling term $\tilde{R}\mathcal{T}$ is significant in one-loop structure of a quantized gravitational theory with torsion \cite{gK2}. (iv) This coupling term modifies the equation of motion of gravitational field and torsion field and results in that the spacetime torsion can propagate in the spacetime even in the absence of spin of matter.
Thus, despite of the possible existence of the dislike ghosts, it is necessary to study the effect of such a coupling on the properties and spacetime structure of a black hole.
For a sake of simplicity,  H. Shabani \textit{et al} \cite{sbh1} considered a special situation in which only the coefficient $a_1$ in $\mathcal{T}$ is taken to be nonzero and the rest of constants are disappeared, and then they obtained some non-trivial static asymptotical flat vacuum solutions, which describe the spacetimes with special structures. These non-trivial solutions are useful for detecting the effects originating from spacetime torsion.

A natural question is whether there exist other asymptotical flat and spherically symmetric solutions in this generalized ECSK theory with the coupling term $\tilde{R}\mathcal{T}$. Here, we will present here a new asymptotical flat and spherically symmetric solution. Our new solution can recover to Schwarzschild solution and Reissner-Nordstr\"{o}m one, which is different from those obtained in Ref. \cite{sbh1}.  Especially, as the parameters take certain special values, this solution can also reduce to the black hole solutions in the braneworld \cite{BHbrane}.
Gravitational lensing is a phenomenon of the deflection of light rays
in the curved spacetime. It is well known that gravitational lensing can provide us
a lot of important signatures about compact objects, which could help to identify black hole and verify alternative theories of gravity in their strong field regime \cite{Ein1,Darwin,KS1,KS2,KS4, VB1,VB2,VB202,Gyulchev,Gyulchev1,Fritt,Bozza1,Eirc1,whisk,
Bhad1,Song1,Song1add,Song101,Song102,Song103,Song2,TSa1,AnAv,bran1,agl1,agl1add,agl1add0,agl101,
gr1,gr1add0,gr1add,gr1noncom,gr2,gr201,gr3,gr4,gr401,gr51,glcom31,glcom32,glcom33,mg1,CSmg,fR1,fR2,
worm1,worm2,worm3,Exdim1,Exdim2,Scatensor,Bardeen,zchen}.
Therefore, in this paper, we also further study the gravitational
lensing in the spacetime described by our new solution.

The paper is organized as follows. In Sec. II, we will firstly present a new asymptotical flat and spherically symmetric solution in the generalized ECKS theory with the coupling term $\tilde{R}\mathcal{T}$. In Sec. III,  we will investigate the propagation of photon in this background and probe the effects of the spacetime parameters on the photon sphere and on the deflection angle for light ray. We also analyze the coefficients in the strong field limit in the cases with photon sphere. Finally, we present a summary.

\section{A black hole solution in the generalized Einstein-Cartan-Kibble-Sciama gravity }

In this section, we will focus on the generalized ECKS theory with the coupling between Ricci scalar $\tilde{R}$ and torsion scalar $\mathcal{T}$, and then present a new black hole solution with torsion. The action in this generalized ECKS theory is \cite{sbh1}
\begin{eqnarray}
S&=&\int d^4x \sqrt{-g}\bigg[-\frac{1}{16\pi G}\bigg(\tilde{R}+\tilde{R}\mathcal{T}\bigg)\bigg],
\label{action}
\end{eqnarray}
with
\begin{eqnarray}
\tilde{R}&=& R+\frac{1}{4}Q_{\alpha\beta\gamma}Q^{\alpha\beta\gamma}
+\frac{1}{2}Q_{\alpha\beta\gamma}Q^{\beta\alpha\gamma}+
Q^{\alpha\;\beta}_{\;\alpha}Q^{\gamma}_{\;\beta\gamma}+2Q^{\alpha\;\beta}_{\;\alpha;\beta},
\nonumber\\
\mathcal{T}&=&a_1Q_{\alpha\beta\gamma}Q^{\alpha\beta\gamma}
+a_2Q_{\alpha\beta\gamma}Q^{\alpha\gamma\beta}+
a_3Q^{\alpha\;\beta}_{\;\alpha}Q^{\gamma}_{\;\gamma\beta}.
\end{eqnarray}
The quantities $\tilde{R}$ and $R$ are Riemann curvature scalar associated with general affine connection $\tilde{\Gamma}^{\alpha}_{\;\mu\nu}$ and Levi-Civita Christoffel connection $\Gamma^{\alpha}_{\;\mu\nu}$, respectively.
The tensor $Q^{\alpha}_{\;\mu\nu}$ describes torsion of spacetime, which is defined by
\begin{eqnarray}
Q^{\alpha}_{\;\mu\nu}=\tilde{\Gamma}^{\alpha}_{\;\mu\nu}-\tilde{\Gamma}^{\alpha}_{\;\nu\mu}.
\end{eqnarray}
It is the antisymmetric part of the general affine connection $\tilde{\Gamma}^{\alpha}_{\;\mu\nu}$. The affine connection $\tilde{\Gamma}^{\alpha}_{\;\mu\nu}$ is related to the Levi-Civita Christoffel connection $\Gamma^{\alpha}_{\;\mu\nu}$ by
\begin{eqnarray}
\tilde{\Gamma}^{\alpha}_{\;\mu\nu}=\Gamma^{\alpha}_{\;\mu\nu}+K^{\alpha}_{\;\mu\nu},
\end{eqnarray}
where $K^{\alpha}_{\;\mu\nu}$ is the contorsion tensor with a form
\begin{eqnarray}
K^{\alpha}_{\;\mu\nu}=\frac{1}{2}\bigg[Q^{\alpha}_{\;\mu\nu}-Q^{\;\alpha}_{\mu\;\nu}
-Q^{\;\alpha}_{\nu\;\mu}\bigg].
\end{eqnarray}
The torsion scalar $\mathcal{T}$ with arbitrary coefficients $a_1$, $a_2$, and $a_3$ is
introduced in New General Relativity by Hayashi and Shirafuji \cite{TNG1}. In particular, the torsion scalar denoted by $\mathcal{T}$ is given by $(a_1, a_2, a_3)=(1/4, 1/2, -1)$ for ``teleparallel equivalent of general relativity (TEGR)". The choices of the coefficients have been explored in the literature
\cite{TNG1,TNG2,TNG3,TNG4,TNG5,TNG6}. The models with $\mathcal{T}$  have been also investigated extensively in cosmology to explain the accelerating expansion of Universe.
Since the coupling $\tilde{R}\mathcal{T}$ could exist in the strong gravity region near a black hole and this coupling is significant in one-loop structure of a quantized gravitational theory with torsion \cite{gK2}, it is necessary to study the effects of interaction term $\tilde{R}\mathcal{T}$ on the properties and spacetime structure of a black hole with torsion.

As in Ref.\cite{sbh1}, we focus on a static spherically symmetric vacuum solution with the metric
\begin{eqnarray}
ds^2 = -Hdt^2 + \frac{dr^2}{F} + r^2 (d\theta^2
+\sin^2\theta d\phi^2), \label{metr}
\end{eqnarray}
where $H$, and $F$ are only functions of the polar coordinate $r$. In this spacetime (\ref{metr}), there are a time-like Killing vector $\xi^{\alpha}_t=(1,0,0,0)$ and a space-like Killing vector $\xi^{\alpha}_s=(0,0,0,1)$ for the metric field $g_{\mu\nu}$, which satisfy the Killing equation $\mathcal{L}_{\xi}g_{\mu\nu}=0$. These two Killing vectors associate with the symmetry under time $t$ displacement and the rotation symmetry around $z-$axis, respectively. Although, in general, the torsion need not has the same symmetries as in the metric field, it is suitable that for a static spherically symmetric solution the torsion is required to be independent of the coordinates $t$ and $\phi$, and then it owns the symmetry under the time $t$ displacement and the rotation symmetry around $z-$axis, which implies that the Lie derivative of torsion $Q^{\alpha}_{\;\mu\nu}$ along the Killing vectors $\xi^{\alpha}_t=(1,0,0,0)$ and  $\xi^{\alpha}_s=(0,0,0,1)$ must be zero. It is easy to find that
the non-vanishing components of the torsion tensor $Q^{\alpha}_{\;\mu\nu}$
\begin{eqnarray}\label{torsion1}
Q^{r}_{\;tr}=-Q^{r}_{\;rt}=A,\,\,\,\,\,\,\,\,\,\,\,
Q^{\theta}_{\;t\theta}=-Q^{\theta}_{\;\theta t}=B,\,\,\,\,\,\,\,\,\,\,\,Q^{\phi}_{\;t\phi}=-Q^{\phi}_{\;\phi t}=B,
\end{eqnarray}
satisfied the Killing equation $\mathcal{L}_{\xi}Q^{\alpha}_{\;\mu\nu}=0$ for both above Killing vector fields $\xi^{\alpha}_t$ and $\xi^{\alpha}_s$. Here $A$ and $B$ depend only on the polar coordinate $r$. The constraint for the torsion $\mathcal{L}_{\xi}Q^{\alpha}_{\;\mu\nu}=0$ has also been applied in cosmological setting \cite{tensorcos1}. Inserting Eqs.(\ref{metr}) and (\ref{torsion1}) into the action (\ref{action}), we can find that the total Lagrangian has a form
\begin{eqnarray}\label{lglri}
\tilde{\mathcal{L}}&=&\frac{\sin\theta}{2H^3}
\sqrt{\frac{H}{F}}\bigg[H-(2a_1+a_2)(A^2+2B^2)+a_3(A+2B)^2\bigg]\bigg[r^2F(2HH''-H'^2)
+HH'(r^2F'+4rF)\nonumber\\&&+4H^2(rF'+F-1)+4r^2HB(2A+B)\bigg],
\end{eqnarray}
where the prime denotes the derivation with respect to $r$.
Making use of Euler-Lagrange equation, we can obtain four differential equations
\begin{eqnarray}\label{eqs1}
\mathcal{Q}_H&=&\bigg[H-2a_1(A^2+2B^2)+a_3(A+2B)^2\bigg]\bigg[2H^2(rF'+F-1)+6r^2HB(2A+B)
+2r^2F(4HH''\nonumber\\&&-5H'^2)+2rHH'(rF'+4F)\bigg]
-rH\bigg[H(rF'+4F)-4rFH'\bigg]\bigg[H-2a_1(A^2+2B^2)+a_3(A+2B)^2\bigg]'
\nonumber\\&&-2r^2FH^2\bigg[H-2a_1(A^2+2B^2)+a_3(A+2B)^2\bigg]''\nonumber\\&&+
\frac{H\mathcal{Q}_A-8r^2BH[H-(2a_1+a_2)(A^2+2B^2)+a_3(A+2B)^2]}{2[(2a_1+a_2)A-a_3(A+2B)]}=0,
\end{eqnarray}
\begin{eqnarray}
\label{eqs2}
\mathcal{Q}_F&\equiv& \bigg\{2H\bigg[(F-1)H+r^2B(2A+B)\bigg]-rFH'(rH'+2H)\bigg\}\bigg[H-2a_1(A^2+2B^2)+a_3(A+2B)^2\bigg]\nonumber\\&+&
rFH(4H+rH')\bigg\{H'-\bigg[(2a_1+a_2)(A^2+2B^2)-a_3(A+2B)^2\bigg]'\bigg\}=0,
\\
\label{eqs3}
\mathcal{Q}_A&\equiv&4r^2BH\bigg[H-(2a_1+a_2)(A^2+2B^2)+a_3(A+2B)^2\bigg]-
\bigg[(2a_1+a_2)A-a_3(A+2B)\bigg]\times\nonumber\\&&\bigg[4r^2BH(2A+B)+4H^2 (rF'+F-1)-r^2FH'^2+rHH'(rF'+4F)+2r^2FHH''\bigg]=0,
\end{eqnarray}
and
\begin{eqnarray}
\label{eqs4}
\mathcal{Q}_B&\equiv&2r^2H(A+B)\bigg[H-(2a_1+a_2)(A^2+2B^2)+a_3(A+2B)^2\bigg]
-\bigg[(2a_1+a_2)B-a_3(A+2B)\bigg]\times\nonumber\\&&\bigg[4r^2BH(2A+B)+4H^2 (rF'+F-1)-r^2FH'^2+rHH'(rF'+4F)+2r^2FHH''\bigg]=0.
\end{eqnarray}
Comparing Eq.(\ref{eqs3}) with Eq.(\ref{eqs4}), we find
\begin{eqnarray}\label{eqs3n}
&&\bigg[(2a_1+a_2)B-a_3(A+2B)\bigg]\mathcal{Q}_A-\bigg[(2a_1+a_2)A-a_3(A+2B)\bigg]\mathcal{Q}_B
\nonumber\\&=&-2(2a_1+a_2-a_3)r^2H(A-B)(A+2B)[H-(2a_1+a_2)(A^2+2B^2)+a_3(A+2B)^2],
\end{eqnarray}
For the equation (\ref{eqs2}), it is obvious that there is a solution $H=(2a_1+a_2)(A^2+2B^2)-a_3(A+2B)^2$.
Inserting $H=(2a_1+a_2)(A^2+2B^2)-a_3(A+2B)^2$ into above equations, we find that the polynomials $\mathcal{Q}_H$,  $\mathcal{Q}_A$  and $\mathcal{Q}_B$ are related by
\begin{eqnarray}
\mathcal{Q}_H=\frac{H\mathcal{Q}_A}{2[(2a_1+a_2)A-a_3(A+2B)]}
=\frac{H\mathcal{Q}_B}{2[(2a_1+a_2)B-a_3(A+2B)]},
\end{eqnarray}
which means that as $H=2a_1(A^2+2B^2)-a_3(A+2B)^2$ equations (\ref{eqs1}), (\ref{eqs3}) and (\ref{eqs4}) can reduce to a single differential equation. This implies that there exist many solutions since there are three independent quantities in the four variables $F$, $H$, $A$ and $B$, but with only an independent differential equation in this case. In order to find a solution, we set $A=-\frac{1}{2}B$ as in Ref.\cite{sbh1}, and then the independent differential equation can be expressed further as
\begin{eqnarray}
\label{eqs5}
2r^2FB''+r(rF'+4F)B'+2(rF'+F-1)B=0.
\end{eqnarray}
It agrees with Eq.(33) in Ref.\cite{sbh1} where $a_2=a_3=0$. In Ref.\cite{sbh1},  this equation is obtained directly by substituting the ansatz (\ref{torsion1}) into the equation of motion for torsion field from the variation of the original action (\ref{action}), and here we get it from the reduced Lagrangian (\ref{lglri}). This means that the truncation of torsion field (\ref{torsion1}) is a consistent truncation. The differential equation (\ref{eqs5}) also implies that once $B$ is selected as a proper form, we can obtain the function $F$ by solving this equation. Letting $B=\sqrt{1-\frac{2m}{r}}$ and rescaling $\frac{9(2a_1+a_2-a_3)}{4}t\rightarrow t$, we can obtain the solution
\begin{eqnarray}\label{solutions0}
H=1-\frac{2m}{r},\,\,\,\,\,\,\,\,\,\,\,\,\,F=\frac{(r-2m)(2r+\alpha)}{r(2r-3m)}, \end{eqnarray}
where $\alpha$ is an integral constant. If taking $F=1-\frac{2\gamma m}{r}$ and rescaling $\frac{9(2a_1+a_2-a_3)}{4}t\rightarrow t$,  we can get the solution
\begin{eqnarray}\label{solutions1}
H=\bigg(1-\frac{1}{\gamma}+\frac{1}{\gamma}\sqrt{1-\frac{2\gamma m}{r}}\bigg)^2,\,\,\,\,\,\,\,\,F=1-\frac{2\gamma m}{r}.
\end{eqnarray}
Interestingly, these two solutions (\ref{solutions0}) and (\ref{solutions1}), respectively, have the same forms as the black hole solutions $I$ and $II$ in the braneworld theory \cite{BHbrane}. This implies that there exists a certain unknown connection between the generalized ECKS theory of gravity (\ref{action}) and the braneworld gravity \cite{BHbrane}.
Here, we set the function $F=1-\frac{2\gamma m}{r}+\frac{q^2}{r^2}$ and rescaling $\frac{9(2a_1+a_2-a_3)}{4}t\rightarrow t$, we can obtain a new black hole solution in the generalized ECKS theory of gravity (\ref{action})
\begin{eqnarray}
\label{sol1}
F&=&1-\frac{2 \gamma m}{r}+\frac{q^2}{r^2},
\nonumber\\
H&=&\frac{1}{(\gamma^2 m^2-q^2)^2}\bigg[\gamma (\gamma-1)m^2+\frac{(1-\gamma)q^2m+(\gamma m^2-q^2)\sqrt{r^2-2 \gamma m r+q^2}}{r}\bigg]^2,
\end{eqnarray}
where $m$, $q$, and $\gamma$ are constants.
This solution is asymptotically flat since both of functions $F$ and $H$ tend to $1$ as $r$ approaches to spatial infinity.
As $\gamma=1$, we find that the solution (\ref{sol1}) reduces to the Reissner-Nordstr\"{o}m type. As $q=0$, it becomes the black hole solution (\ref{solutions1}). Moreover, we also note that the metric form  (\ref{sol1}) can reduce to that of a scalar-tensor wormhole obtained in Ref. \cite{wormRS} with the parameter replacement $q^2\rightarrow-\beta$, $\gamma m\rightarrow m$ and $(\gamma m^2-q^2)/(\gamma(\gamma-1)m^2)\rightarrow \eta$.
As in the case of the black hole solutions $I$ and $II$ in the braneworld \cite{BHbrane}, the solution (\ref{sol1}) will be expressed in terms of the ADM mass $m$ and the parameterized post-Newtonian (PPN) parameter $\beta_0=\frac{\gamma+1}{2}$. The parameter $q$ is similar to a ``tidal-like charge". The position of outer event horizon lies at
\begin{eqnarray}\label{EH1}
r_H=\gamma m+\sqrt{\gamma^2m^2-q^2},
\end{eqnarray}
which is defined by equation $F=0$. Especially, the  surface gravity constant of event horizon for the black hole (\ref{sol1}) is
\begin{eqnarray}
\kappa=\frac{1}{2}\sqrt{\frac{g^{rr}}{-g_{tt}}}\frac{-dg_{tt}}{dr}\bigg|_{r=r_H}=
\frac{1}{2}\sqrt{\frac{F(r)}{H(r)}}\frac{d H(r)}{dr}\bigg|_{r=r_H}=\frac{\gamma m^2-q^2}{\sqrt{\gamma^2m^2-q^2}(\gamma m+\sqrt{\gamma^2m^2-q^2})^2}.
\end{eqnarray}
Thus, the presence of parameter $\gamma$ will bring some particular spacetime properties differed from those in the Einstein's general relativity, which could make a great deal influence on the propagation of photon in the spacetime with metric functions (\ref{sol1}).

\section{The deflection angle for light ray and strong gravitational lensing in the spacetime with torsion}

Let us to study the deflection angle for light ray and the corresponding
gravitational lensing in the background of a black hole spacetime with torsion (\ref{sol1}).
In the spacetimes with non-vanishing torsion, it is well known that there are two types of preferred curves, i.e., the geodesic and the autoparallel curves \cite{ges1,ges2}. The former are curves of extremal length and the corresponding line element between
any two points depends only on the metric. In other words, geodesics is a global concept defined as the shortest path between two points.
Thus, the geodesic equation is \cite{ges1,ges2}
\begin{eqnarray}
\frac{d^2x^{\mu}}{d\lambda^2}+\Gamma^{\mu}_{\nu\tau}\frac{dx^{\nu}}{d\lambda}
\frac{dx^{\tau}}{d\lambda}=0,
\end{eqnarray}
which is the same as in the associated Riemannian spacetime. In the spacetime (\ref{sol1}), the non-zero components of the geodesic equation are
\begin{eqnarray}\label{ges1}
&&\frac{d^2t}{d\lambda^2}+\frac{H'}{H}\frac{dt}{d\lambda}
\frac{dr}{d\lambda}=0,\nonumber\\
&&\frac{d^2r}{d\lambda^2}+\frac{1}{2}FH'\bigg(\frac{dt}{d\lambda}\bigg)^2
-\frac{F'}{2F}\bigg(\frac{dr}{d\lambda}\bigg)^2-Fr\bigg(\frac{d\theta}{d\lambda}\bigg)^2
-Fr\sin\theta\bigg(\frac{d\phi}{d\lambda}\bigg)^2=0,\nonumber\\
&&\frac{d^2\theta}{d\lambda^2}+\frac{2}{r}\frac{dr}{d\lambda}
\frac{d\theta}{d\lambda}-\sin\theta\cos\theta\bigg(\frac{d\phi}{d\lambda}\bigg)^2=0,\nonumber\\
&&\frac{d^2\phi}{d\lambda^2}+\frac{2}{r}\frac{dr}{d\lambda}
\frac{d\phi}{d\lambda}+\frac{2\cos\theta}{\sin\theta}\frac{d\theta}{d\lambda}\frac{d\phi}{d\lambda}=0.
\end{eqnarray}
Autoparallel curves are the lines along which the
covariant derivative of the velocity ($u^a=dx^a/d\lambda$) satisfies $u^b\nabla_{b}u^a=0$, which are defined locally as the straightest path in spacetime \cite{ges1,ges2}. The autoparallel equation has a form \cite{ges1,ges2}
\begin{eqnarray}\label{ges20}
\frac{d^2x^{\mu}}{d\lambda^2}+\Gamma^{\mu}_{\nu\tau}\frac{dx^{\nu}}{d\lambda}
\frac{dx^{\tau}}{d\lambda}+Q_{(\nu\tau)}^{\;\;\;\;\;\mu}\frac{dx^{\nu}}{d\lambda}
\frac{dx^{\tau}}{d\lambda}=0,
\end{eqnarray}
where $Q_{(\nu\tau)}^{\;\;\;\;\;\mu}=\frac{1}{2}(Q_{\nu\tau}^{\;\;\;\mu}
+Q_{\tau\nu}^{\;\;\;\mu})$.
In the background (\ref{sol1}), equation of motion for a point particle moving along
autoparallel trajectory  can be expressed as
\begin{eqnarray}\label{ges2}
&&\frac{d^2t}{d\lambda^2}+\frac{H'}{H}\frac{dt}{d\lambda}
\frac{dr}{d\lambda}-\frac{1}{2\sqrt{H}F}\bigg(\frac{dr}{d\lambda}\bigg)^2+
\frac{r^2}{\sqrt{H}}\bigg(\frac{d\theta}{d\lambda}\bigg)^2+\frac{r^2\sin^2\theta }{\sqrt{H}}\bigg(\frac{d\phi}{d\lambda}\bigg)^2=0,\nonumber\\
&&\frac{d^2r}{d\lambda^2}+\frac{1}{2}FH'\bigg(\frac{dt}{d\lambda}\bigg)^2
-\frac{F'}{2F}\bigg(\frac{dr}{d\lambda}\bigg)^2-Fr\bigg(\frac{d\theta}{d\lambda}\bigg)^2
-Fr\sin\theta\bigg(\frac{d\phi}{d\lambda}\bigg)^2-\frac{\sqrt{H}}{4}
\frac{dr}{d\lambda}\frac{dt}{d\lambda}=0,\nonumber\\
&&\frac{d^2\theta}{d\lambda^2}+\frac{2}{r}\frac{dr}{d\lambda}
\frac{d\theta}{d\lambda}-\sin\theta\cos\theta\bigg(\frac{d\phi}{d\lambda}\bigg)^2+\frac{\sqrt{H}}{2}
\frac{d\theta}{d\lambda}\frac{dt}{d\lambda}=0,\nonumber\\
&&\frac{d^2\phi}{d\lambda^2}+\frac{2}{r}\frac{dr}{d\lambda}
\frac{d\phi}{d\lambda}
+\frac{2\cos\theta}{\sin\theta}\frac{d\theta}{d\lambda}\frac{d\phi}{d\lambda}
+\frac{\sqrt{H}}{2}
\frac{d\phi}{d\lambda}\frac{dt}{d\lambda}=0.
\end{eqnarray}
Obviously, in the spacetime (\ref{sol1}), the motion of particle along the autoparallel curve (\ref{ges2}) differs from that along the geodesic curve (\ref{ges1}). Due to the presence of the torsion-coupling term in Eq.(\ref{ges20}), the four component equations (\ref{ges2}) of autoparallel curve are coupled together so that they are not variable-separable, which means that the motion of photon along the autoparallel curve becomes non-integrable so that there exists irregular orbit for photon. In order to obtain these irregular orbits, we must resort to the complicated numerical techniques.  However, the geodesic motion of photon (\ref{ges1}) is integrable and all of orbits are regular.
Moreover, in order to compare with the cases without torsion, here we focus only on the motion of the photon along geodesic curve in the spacetime (\ref{sol1}).

Considered that the nature of the spherically symmetric spacetime, we here consider only the case where both the source and the observer lie in the equatorial plane so that the orbit of the photon is limited on the same plane in the background spacetime.
With the condition $\theta=\pi/2$, the metric (\ref{sol1}) can be expressed as
\begin{eqnarray}
ds^2=-\mathcal{A}(r)dt^2+\mathcal{B}(r)dr^2+C(r)d\phi^2,\label{grm}
\end{eqnarray}
with
\begin{eqnarray}
\mathcal{A}(r)=H, \;\;\;\;\mathcal{B}(r)&=&1/F,\;\;\;\; C(r)=r^2.
\end{eqnarray}
The geodesics for the photon (\ref{ges1}) in the spacetime (\ref{grm}) obey
\begin{eqnarray}
\frac{dt}{d\lambda}&=&\frac{1}{\mathcal{A}(r)},\label{u3}\\
\frac{d\phi}{d\lambda}&=&\frac{J}{C(r)},\label{u4}\\
\bigg(\frac{dr}{d\lambda}\bigg)^2&=&\frac{1}{\mathcal{B}(r)}\bigg[\frac{1}{\mathcal{A}(r)}
-\frac{J^2}{C(r)}\bigg].\label{cedi}
\end{eqnarray}
where $J$ is the angular momentum of the photon and $\lambda$ is an affine parameter along the null geodesics. Here, the energy of photon is set to $E=1$. For a black hole spacetime (\ref{sol1}), the impact parameter $u(r_0)$ for a photon can be expressed as
\begin{eqnarray}
u(r_0)=J(r_0)=\frac{r_0}{\sqrt{H(r_0)}},
\end{eqnarray}
where $r_0$ is the closest distance between the photon and black hole. It is well known that the photon sphere plays an important role in the propagation of photon in a curved spacetime. In the background of a black hole spacetime (\ref{sol1}), the radius of the photon sphere $r_{ps}$ is the largest real root of the equation
\begin{eqnarray}
&&\mathcal{A}(r)C'(r)-\mathcal{A}'(r)C(r)=2\bigg[m(\gamma-1)(q^2-\gamma mr)+(q^2-\gamma m^2)\sqrt{r^2-2 \gamma m r+q^2}\bigg]\nonumber\\&\times&
\bigg[(\gamma m^2-q^2)(r^2-3\gamma mr+2q^2)+m(\gamma-1)(\gamma mr-2q^2)\sqrt{r^2-2 \gamma m r+q^2}\bigg]=0.\label{root}
\end{eqnarray}
\begin{figure}
\begin{center}
\includegraphics[width=6cm]{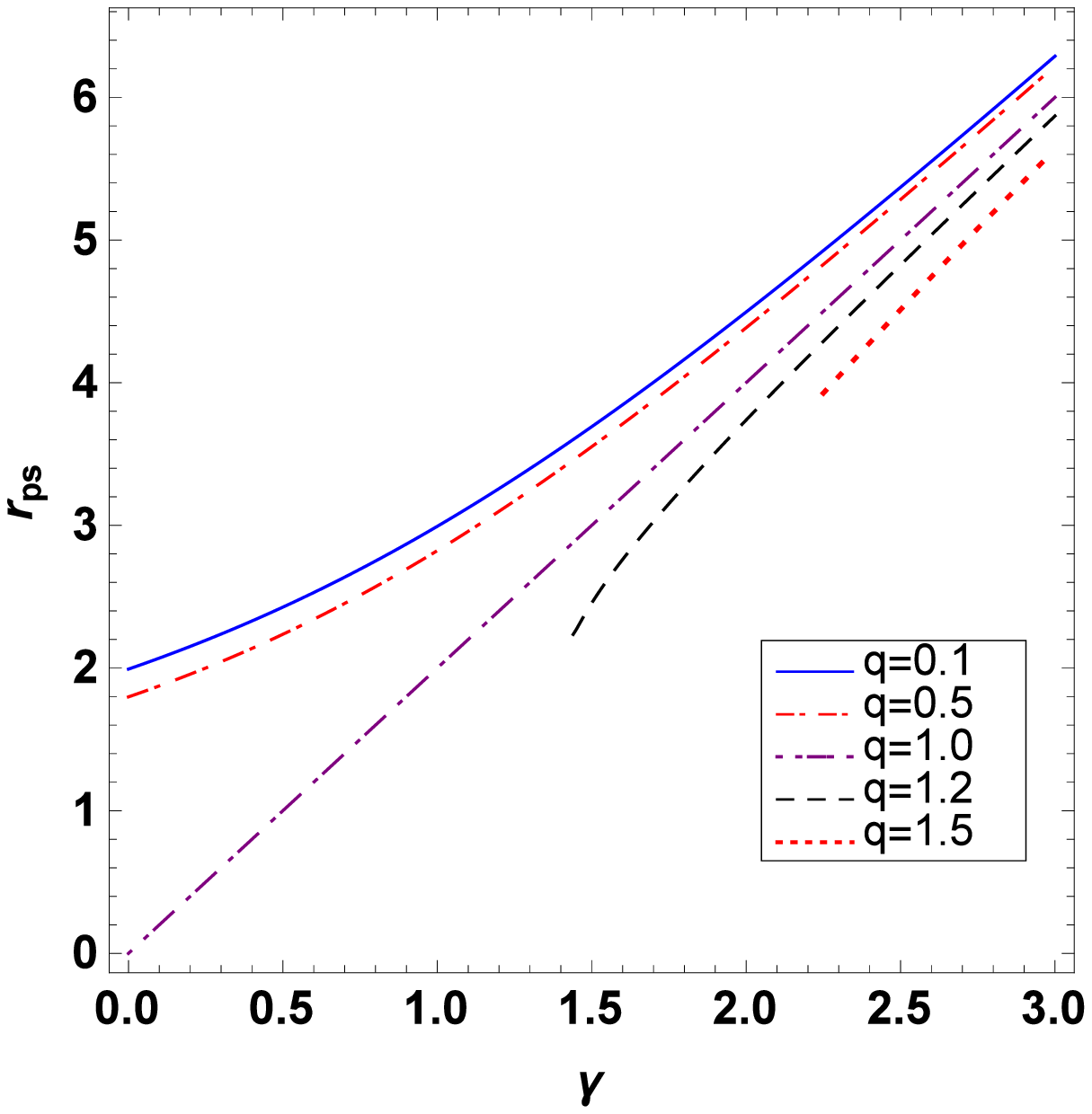}\includegraphics[width=6cm]{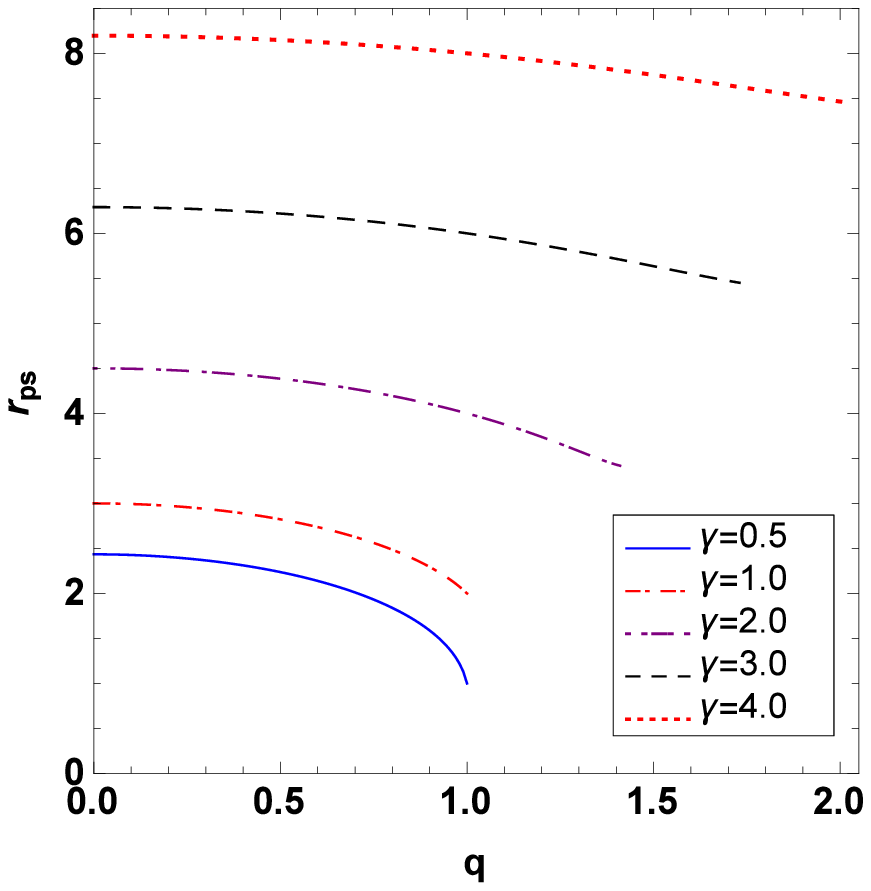}
\caption{Change of the photon sphere radius $r_{ps}$ with parameters $q$ and $\gamma$ in the background of a black hole spacetime with torsion (\ref{sol1}). Here, we set $m=1$.}
\end{center}
\end{figure}
As expected, the radius of the photon sphere $r_{ps}$ depends on both the parameters $q$ and $\gamma$ of the black hole. However, the appearance of $\gamma$ yields that the form of the photon sphere radius become very complicated in this case. In Fig.1, we present the variety of the photon sphere $r_{ps}$  with the parameter $q$ and $\gamma$ of by solving Eq. (\ref{root}) numerically. It is shown that the photon sphere radius increases with
the parameter $\gamma$ and decreases with $q$. The change of $r_{ps}$ with $q$ is similar to that in the Reissner-Nordstr\"{o}m spacetime.
\begin{figure}
\begin{center}
\includegraphics[width=6cm]{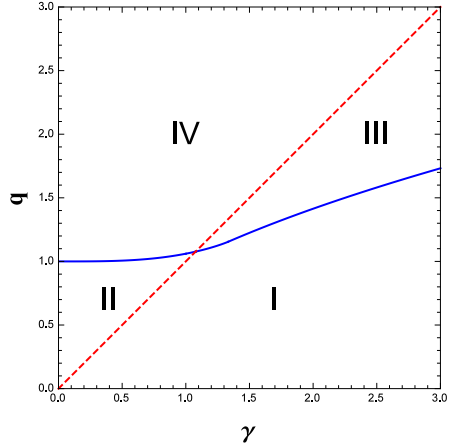}
\caption{ The boundary of the existence of horizon (red dashed line) and of the photon sphere (blue line) in the background of a black hole spacetime with torsion (\ref{sol1}). Here, we set $m=1$.}
\end{center}
\end{figure}
Moreover, we find that the photon sphere $r_{ps}$ exists only in the regime $q<q_c$. The value of the upper limit $q_c$ depends
on the parameter $\gamma$, and its form can be expressed as
\begin{eqnarray}\label{phcri}
q_c=\left\{\begin{array}{lll}
\frac{m}{2\sqrt{6}}\bigg[\frac{\mathcal{W}^{2/3}+2(9\gamma^2+4)\mathcal{W}^{1/3}+
(9\gamma^2-8)^2}{\mathcal{W}^{1/3}}\bigg]^{1/2},&&0<\gamma\leq1,\\\\
\frac{m}{4\sqrt{3}}\bigg[\frac{(-1-\sqrt{3}i)\mathcal{W}^{2/3}+4(9\gamma^2+4)\mathcal{W}^{1/3}
+(-1+\sqrt{3}i)(81\gamma^4+64)^2}{\mathcal{W}^{1/3}}\bigg]^{1/2},&&1\leq\gamma\leq \frac{4}{3},\\\\
\sqrt{\gamma},&&\gamma\geq \frac{4}{3},
\end{array}\right.
\end{eqnarray}
where
\begin{eqnarray}
\mathcal{W}=54\gamma^2\sqrt{2(27\gamma^3-18\gamma-8)(3\gamma-4)^3}|\gamma-1| +2187\gamma^6-5832\gamma^5+4860\gamma^4-1728\gamma^2+512.
\end{eqnarray}
Comparing Eq. (\ref{EH1}) with Eq. (\ref{phcri}), one can find that the condition of existence of horizons is not inconsistent with that of the photon sphere, which is also shown in Fig.(2). The red dashed line is the boundary for existence of horizon and  the blue solid line is for photon sphere. It is obvious that
the whole region in the parameter panel $(\gamma, q)$ is split into four regions $I$-$IV$ by those critical curves.
When the parameters $(\gamma, q)$ lie in the region $I$, there exist both horizon and  photon sphere radius $r_{ps}$, which is similar to that of in the static black hole spacetime in the Einstein's general relativity. When $(\gamma, q)$  lie in the region $IV$, there is neither horizon nor photon sphere, which corresponds to case of strong naked singularity where the singularity is completely naked \cite{KS4,Gyulchev1}. When $(\gamma, q)$  is located in region $II$, there exists only photon sphere but no horizon, which corresponds to case of weak naked singularity where the singularity is covered by the photon sphere \cite{KS4,Gyulchev1}.
When $(\gamma, q)$  lies in the region $III$, there is horizon but no photon sphere.
These four situations are similar to those in the black hole spacetime with a torsion \cite{zchen}. The special situation in which the black hole owns horizon but no photon sphere is absent in other theories of gravity. Thus, the presence of torsion changes the spacetime structure which will affect the propagation of photon in the background spacetime.

Let us now to discuss the behavior of the deflection angle of light ray in the spacetime
described by a metric with torsion (\ref{sol1}).  For the photon coming from infinity, the deflection angle in a curved spacetime can be expressed as
\begin{eqnarray}
\alpha(r_0)=I(r_0)-\pi,
\end{eqnarray}
where $r_0$ is the closest approach distance and $I(r_0)$ is \cite{KS1}
\begin{eqnarray}
I(r_0)=2\int^{\infty}_{r_0}\frac{\sqrt{\mathcal{B}(r)}dr}{\sqrt{C(r)}
\sqrt{\frac{C(r)\mathcal{A}(r_0)}{C(r_0)\mathcal{A}(r)}-1}}.\label{int1}
\end{eqnarray}
\begin{figure}[ht]\label{pas22}
\begin{center}
\includegraphics[width=6cm]{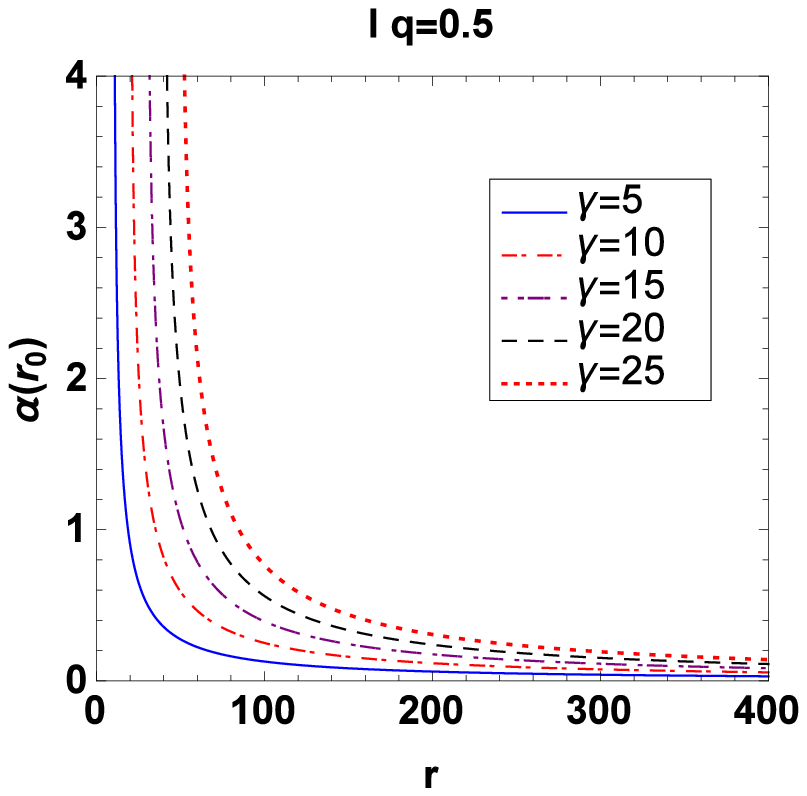}\includegraphics[width=6cm]{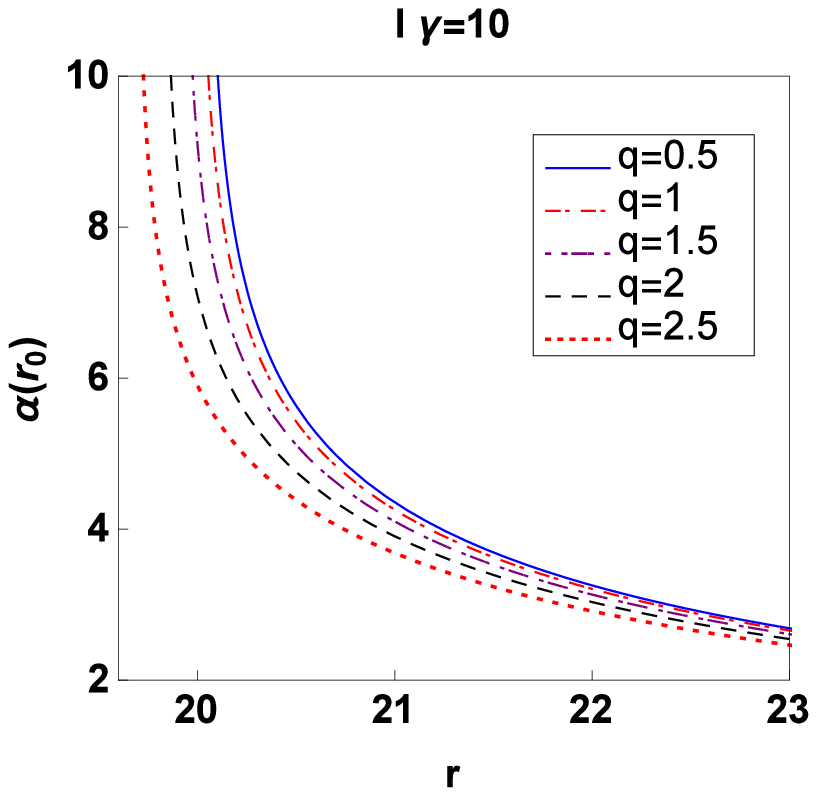}\\
\includegraphics[width=6cm]{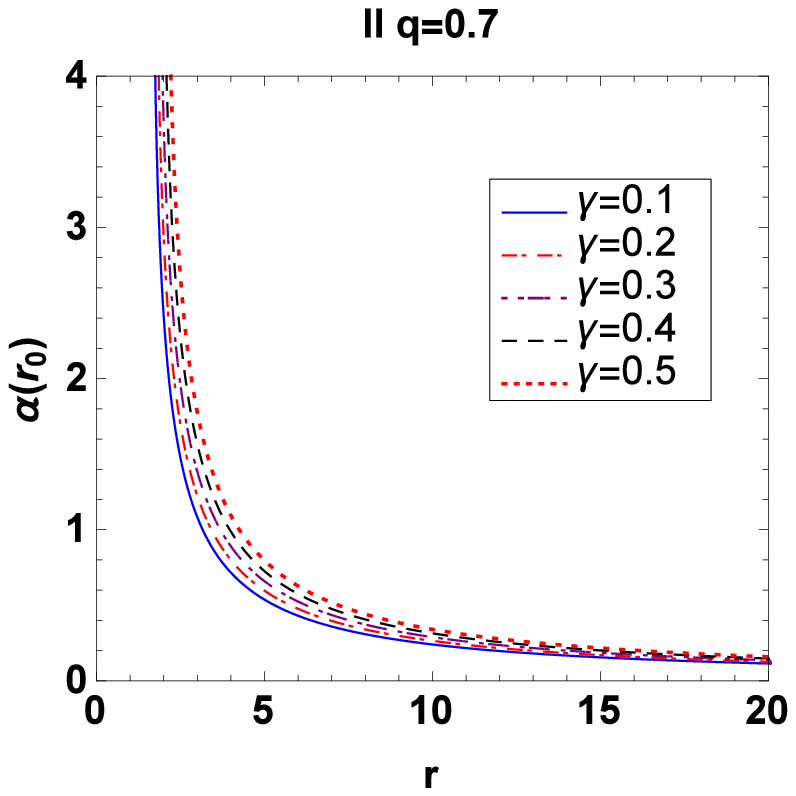}\includegraphics[width=6cm]{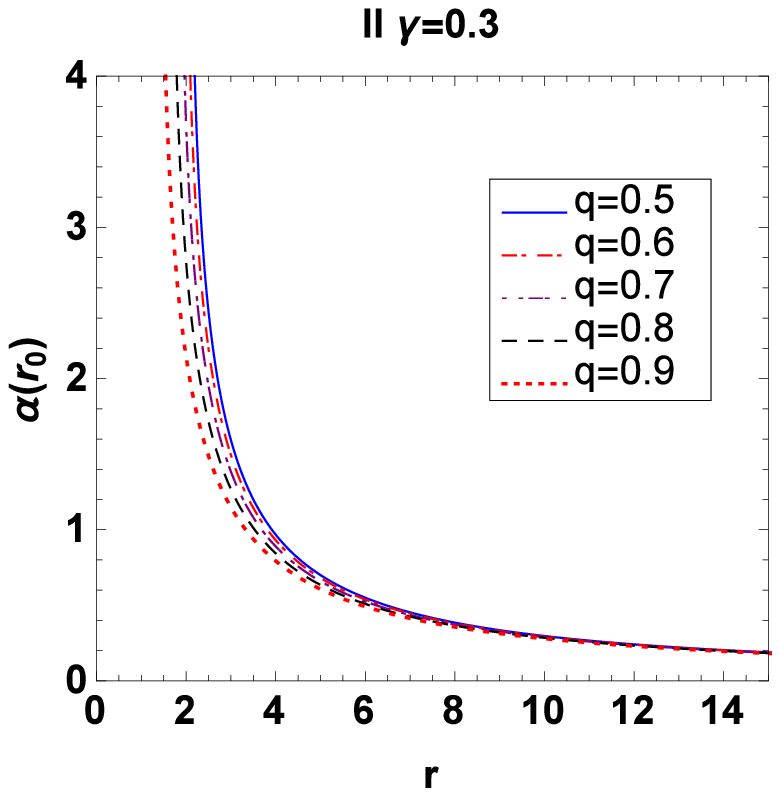}
\caption{Deflection angle $\alpha(r_0)$ as a function of the closest
distance of approach $r_0$ for the cases with photon sphere. The panels in the upper and bottom rows correspond to the cases in which the parameters ($\gamma, q$) are located in the regions $I$ and $II$ in Fig.(2), respectively. Here, we set $m=1$.}
\end{center}
\end{figure}
\begin{figure}[ht]\label{pas3}
\begin{center}
\includegraphics[width=6cm]{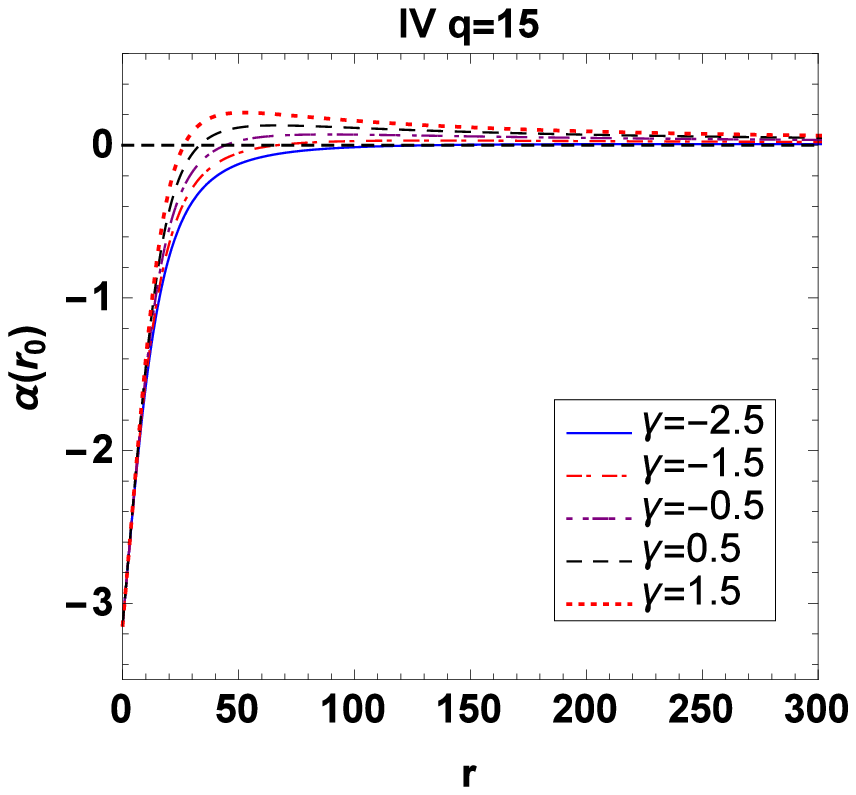}\includegraphics[width=6cm]{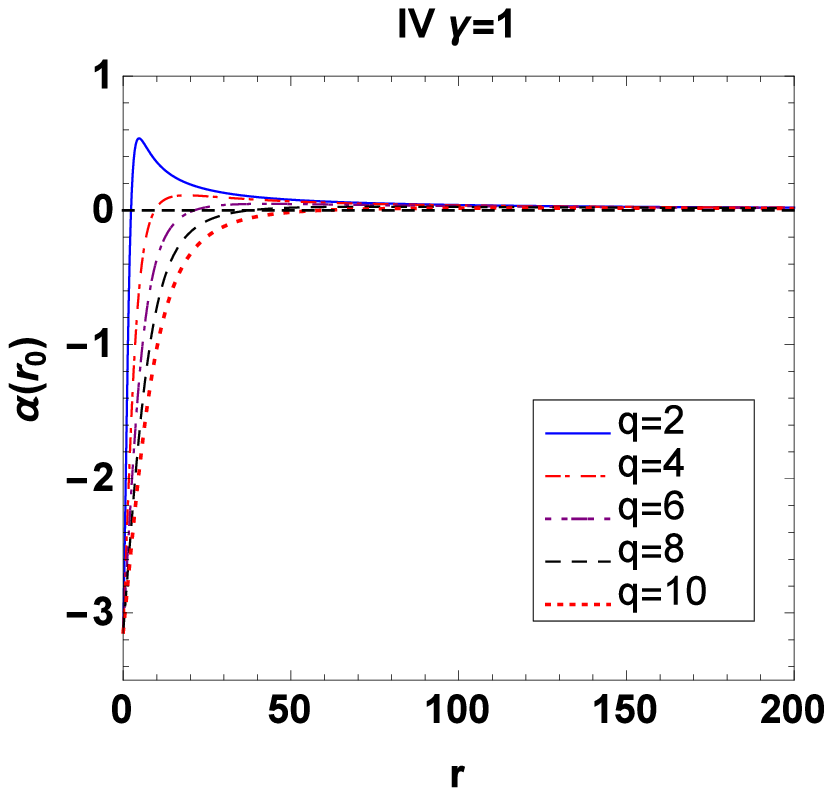}
\caption{Deflection angle $\alpha(r_0)$ as a function of the closest
distance of approach $r_0$ for the case without photon sphere and horizon in which the parameters ($\gamma, q$) lie in the region $IV$ in Fig.(2). Here, we set $m=1$.}
\end{center}
\end{figure}
\begin{figure}[ht]\label{pas4}
\begin{center}
\includegraphics[width=6cm]{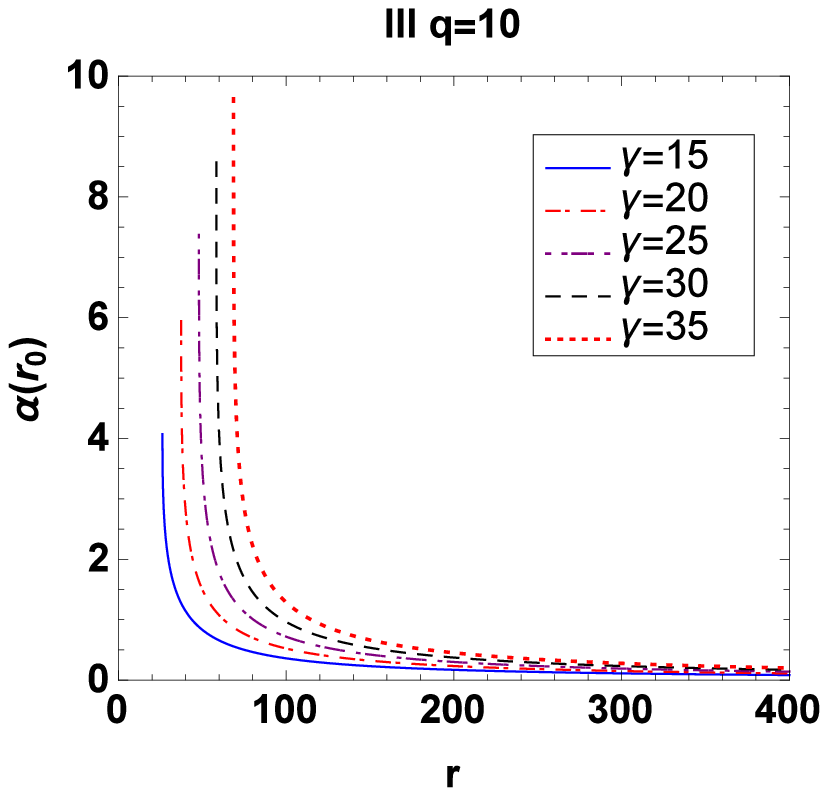}\includegraphics[width=6cm]{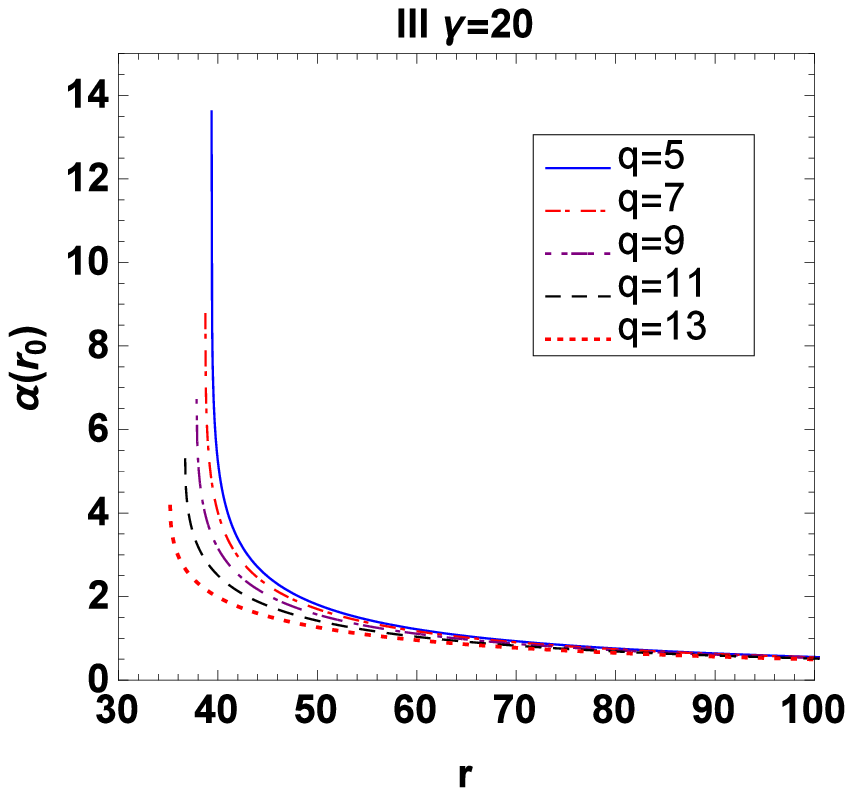}
\caption{Deflection angle $\alpha(r_0)$ as a function of the closest
distance of approach $r_0$ for the case with horizon and no photon sphere in which the parameters ($\gamma, q$) lie in the region $III$ in Fig.(2). Here, we set $m=1$.}
\end{center}
\end{figure}
In Figs.(3)-(5), we plot the change of the deflection angle $\alpha(r_0)$ with the distance of approach $r_0$ for different parameters $\gamma$ and $q$ in the spacetime with torsion (\ref{sol1}). For the cases with  photon sphere, i.e., the parameters ($\gamma, q$)  lie in the region $I$ or $II$ in Fig.(2), the deflection angle for different $\gamma$ and $q$ strictly increases with the decreases
of the closest distance of approach $r_{0}$ and finally becomes infinite as $r_{0}$ tends to the respective photon sphere radius
$r_{ps}$, i.e., $\text{lim}_{r_0\rightarrow r_{ps}}\alpha(r_{0})=\infty$, which is shown in Fig.(3). In Fig.(4), we present the deflection angle in the case in which the parameters ($\gamma, q$)  lie in the region $IV$ in Fig.(2).  There is neither horizon nor photon sphere so that the singularity is naked completely. It is shown that the deflection angle of the light ray closing to the naked singularity tends to a finite value $-\pi$ for different $\gamma$ and $q$, which means that the photon could not be captured by the compact object so that the photon goes back along the original direction in this situation. This behavior can be regarded as a common feature of gravitational lensing by strong naked singularity. As the parameters ($\gamma, q$) lie in the region $III$ in Fig.(2), there exists horizon but no photon sphere, we find that the deflection angle of the light finally becomes a finite value as $r_{0}$ tends to the respective event horizon radius
$r_{H}$, i.e., $\text{lim}_{r_0\rightarrow r_{H}}\alpha(r_{0})=\alpha_{r_H}$. This behavior differs from those in the black hole with a torsion considered in Ref.\cite{zchen} in which the deflection angle of the light finally becomes unlimited large as $r_{0}$ tends to the event horizon radius. It could be understand by a fact that due to nonexistence of photon sphere the photon is captured directly by black hole before it make infinite
complete loops around the central object in this case. Moreover, we find that the deflection angle $\alpha_{r_H}$ increases with the parameter $\gamma$ and decreases with $q$.
In the far-field limit, the deflection angle can be approximated as
\begin{eqnarray}
\alpha|_{r_0\rightarrow\infty}\simeq\frac{2(\gamma+1)m}{r_0}
+\frac{(6\pi-8)(\gamma+1)m^2+3\pi(\gamma^2m^2-q^2)}{4r^2_0},
\end{eqnarray}
which means that $\text{lim}_{r_0\rightarrow\infty}\alpha(r_{0})=0$
for all values of parameters $\gamma$ and $q$, which is a common feature in all asymptotical flat spacetimes.

We are now in position to study the strong gravitational lensing by a compact object (\ref{sol1}) with the photon sphere and then probe how the parameters $\gamma$ and $q$
affect the coefficients in the
strong field limit. Using of the method developed by Bozza \cite{VB1}, one can define a variable
\begin{eqnarray}
z=1-\frac{r_0}{r},
\end{eqnarray}
and then rewrite the integral (\ref{int1}) as
\begin{eqnarray}
I(r_0)=\int^{1}_{0}R(z,r_0)f(z,r_0)dz,\label{in1}
\end{eqnarray}
with
\begin{eqnarray}
R(z,r_0)&=&2\sqrt{\mathcal{A}(r)\mathcal{B}(r)C(r)},
\end{eqnarray}
\begin{eqnarray}
f(z,r_0)&=&\frac{1}{\sqrt{\mathcal{A}(r_0)C(r)-\mathcal{A}(r)C(r_0)}}.
\end{eqnarray}
The function $R(z, r_0)$ is regular for all values of $z$ and $r_0$, but $f(z, r_0)$ diverges as $z$ tends to zero. Thus, one can split
the integral (\ref{in1}) into the divergent part $I_{D_1}(r_0)$ and the regular
part $I_{R}(r_0)$,
\begin{eqnarray}
I_{D}(r_0)&=&\int^{1}_{0}R(0,r_{ps})f_{0}(z,r_0)dz, \nonumber\\
I_{R}(r_0)&=&\int^{1}_{0}[R(z,r_0)f(z,r_0)-R(0,r_{ps})f_{0}(z,r_0)]dz
\label{intbr}.
\end{eqnarray}
Expanding the argument of the square
root in $f(z,r_0)$ to the second order in $z$, one can obtain
\begin{eqnarray}
f_{0}(z,r_0)=\frac{1}{\sqrt{\alpha(r_0)z+\beta(r_0)z^2}},
\end{eqnarray}
with
\begin{eqnarray}
\alpha(r_0)&=& \frac{2}{(\gamma^2-q^2)^2\sqrt{r^2_0-2 \gamma r_0+q^2}}\bigg[(\gamma-1)(q^2-\gamma r_0)+(q^2-\gamma)\sqrt{r^2_0-2 \gamma r_0+q^2}\bigg]\nonumber\\&\times&
\bigg[(\gamma-q^2)(r^2_0-3\gamma r_0+2q^2)+(\gamma-1)(\gamma r_0-2q^2)\sqrt{r^2_0-2 \gamma r_0+q^2}\bigg],  \nonumber\\
\beta(r_0)&=&\frac{1}{(\gamma^2-q^2)^2}\bigg\{(\gamma-1)(q^2-\gamma)
\bigg[\frac{r_0(\gamma r_0-q^2)[2r_0^3-6\gamma r_0^2-2\gamma q^2+3(\gamma^2+q^2)r_0]}{(r^2_0-2 \gamma r_0+q^2)^{3/2}}\nonumber\\&+&\frac{2r_0(\gamma-r_0)(\gamma r_0-2 q^2)}{\sqrt{r^2_0-2 \gamma r_0+q^2}}-2(3\gamma r_0-2q^2)\sqrt{r^2_0-2 \gamma r_0+q^2}\bigg]+2q^6-6\gamma^3r_0\nonumber\\&+&
 [3r_0^2-6\gamma r_0+2(\gamma^2-4\gamma+1)]q^4-2\gamma[3r_0^2+3(\gamma^2-4\gamma+1) r_0-\gamma]q^2+3\gamma^2r_0^2(\gamma^2-2\gamma+2)\bigg\}.
\end{eqnarray}
\begin{figure}[h]\label{pas10}
\begin{center}
\includegraphics[width=6cm]{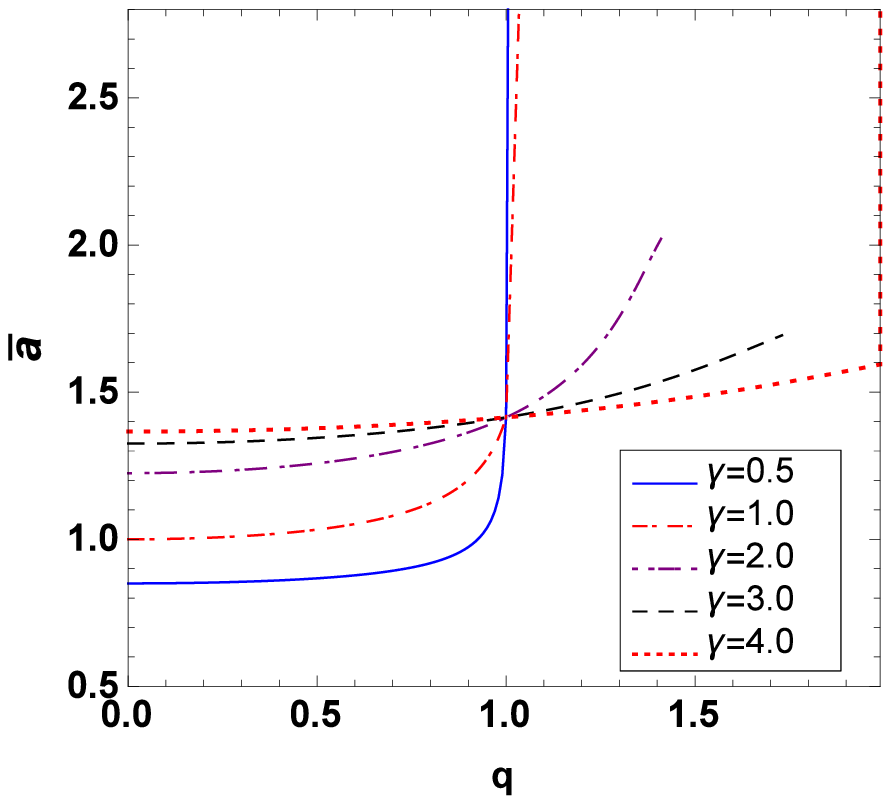}\includegraphics[width=6cm]{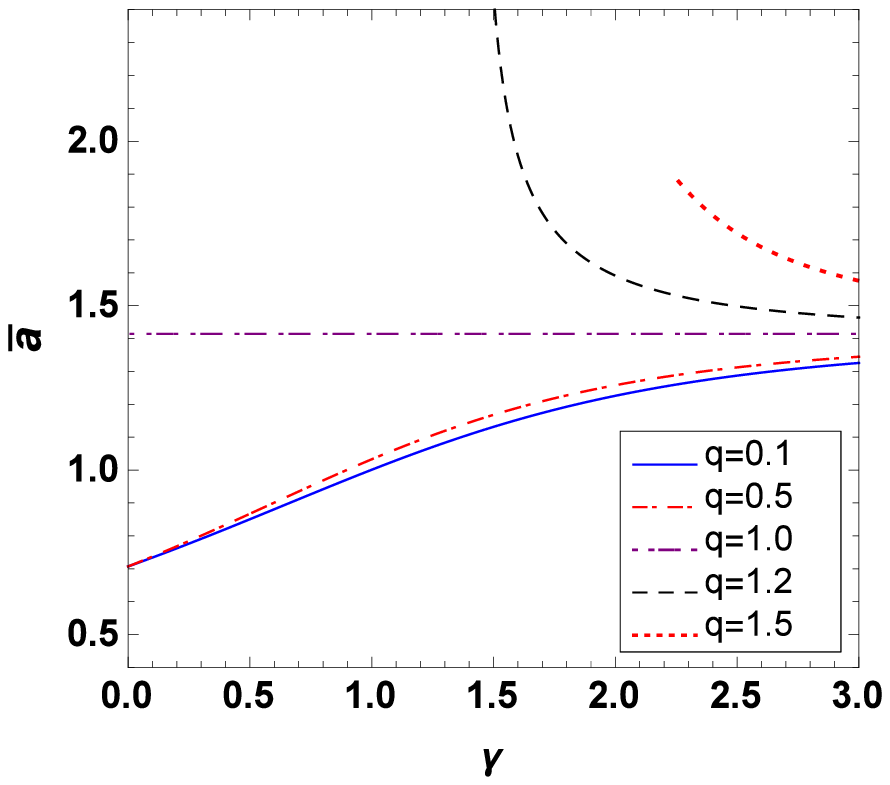}\\
\includegraphics[width=6cm]{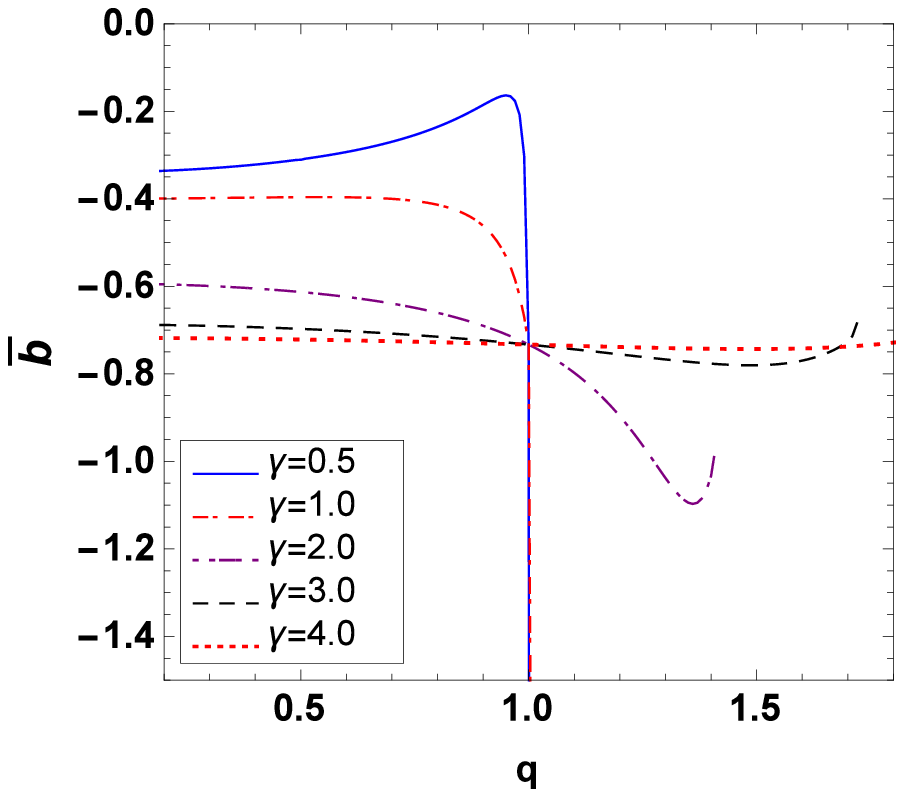}\includegraphics[width=6cm]{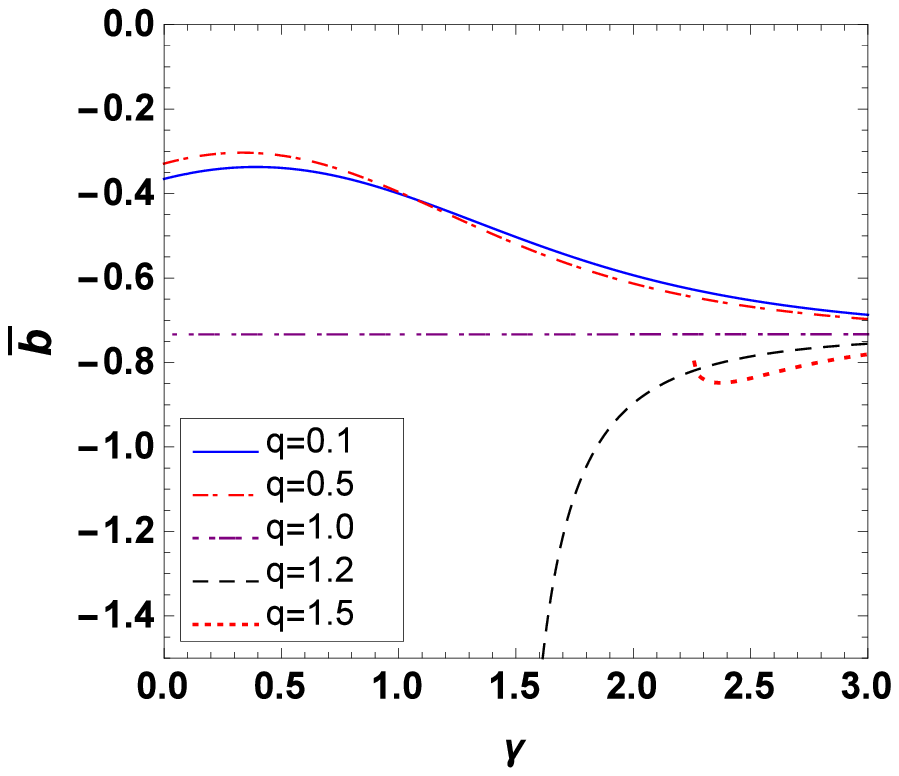}
\caption{Change of the strong deflection limit coefficients $\bar{a}$ and $\bar{b}$ with the parameters $\gamma$ and $q$ in the spacetime with torsion (\ref{sol1}).}
\end{center}
\end{figure}
It is obvious that the
coefficient $\alpha(r_0)$ vanishes as $r_0$ tends to the radius of photon sphere $r_{ps}$, and then the leading term of the
divergence in $f_0(z,r_0)$ is $z^{-1}$. This means that near the photon sphere the deflection angle of light ray can be expressed as \cite{VB1}
\begin{eqnarray}
\alpha(\theta)=-\bar{a}\ln{\bigg(\frac{\theta
D_{OL}}{u_{ps}}-1\bigg)}+\bar{b}+O(u-u_{ps}),
\end{eqnarray}
with
\begin{eqnarray}
&\bar{a}&=\frac{R(0,r_{ps})}{2\sqrt{\beta(r_{ps})}}, \nonumber\\
&\bar{b}&=
-\pi+b_{R}+\bar{a}\ln{\frac{r^2_{ps}[C''(r_{ps})\mathcal{A}(r_{ps})-C(r_{ps})\mathcal{A}''(r_{ps})]}{u_{ps}
\sqrt{\mathcal{A}^3(r_{ps})C(r_{ps})}}}, \nonumber\\
&b_{R}&=I_{R}(r_{ps}), \;\;\;\;\;u_{ps}=\frac{r_{ps}}{\sqrt{\mathcal{A}(r_{ps})}},
\end{eqnarray}
which indicates clearly that the deflection angle
diverges logarithmically where the light is close to the photon sphere. The quantity
$D_{OL}$ is the distance between gravitational
lens object and observer, $\bar{a}$ and $\bar{b}$ are the strong field limit
coefficients which depend on the spacetime functions at the photon sphere.
The changes of the coefficients ($\bar{a}$ and $\bar{b}$ ) with the parameters $\gamma$ and $q$ is shown in Fig.(6). For the fixed $\gamma$, the coefficient $\bar{a}$ increases monotonously with $q$. With increase of $q$, the coefficient $\bar{b}$ first increases and then decreases for the smaller $\gamma$, but first decreases and then increases for the larger $\gamma$. With increase of the parameter $\gamma$, $\bar{a}$ increases monotonously as $q<1$ and decreases as $q>1$, $\bar{b}$ first increases and then decreases as $q<1$. For the case with $q>1$, $\bar{b}$ increases with $\gamma$ as the quantity $q-1$ is small. With the further increase of $q$, the change of $\bar{b}$ shows gradually a tendency of first decreasing and then increasing.

Let us now to estimate the numerical values for the observables of gravitational lensing  in the strong field limit by assuming that the spacetime of the supermassive compact object
at the Galactic center of Milky Way can be described by the metric (\ref{sol1}).
\begin{figure}[h]\label{th10}
\begin{center}
\includegraphics[width=6cm]{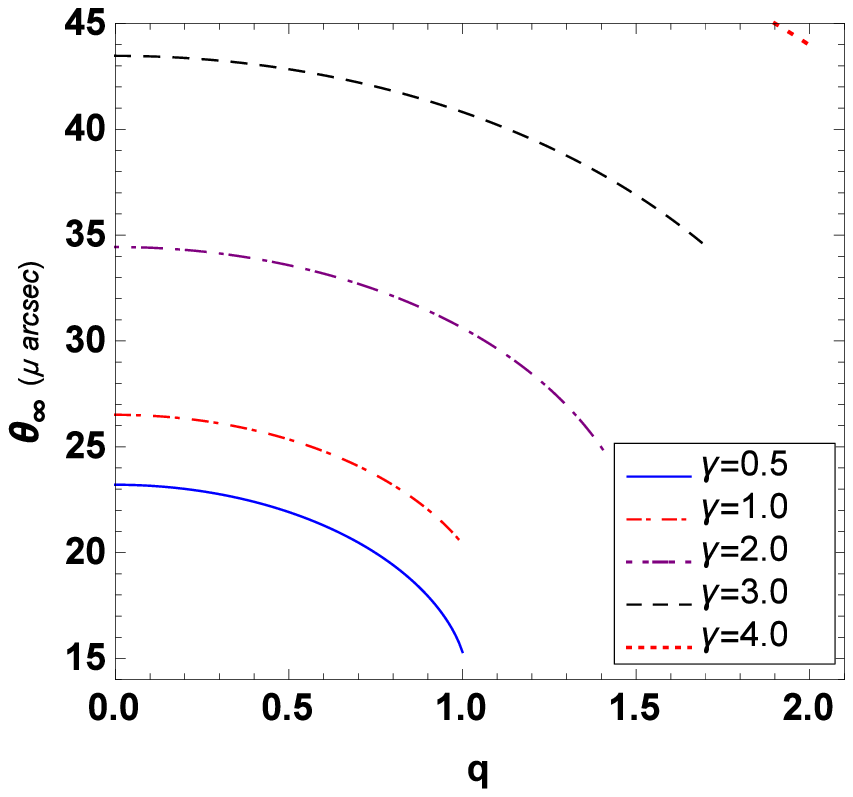}\includegraphics[width=6cm]{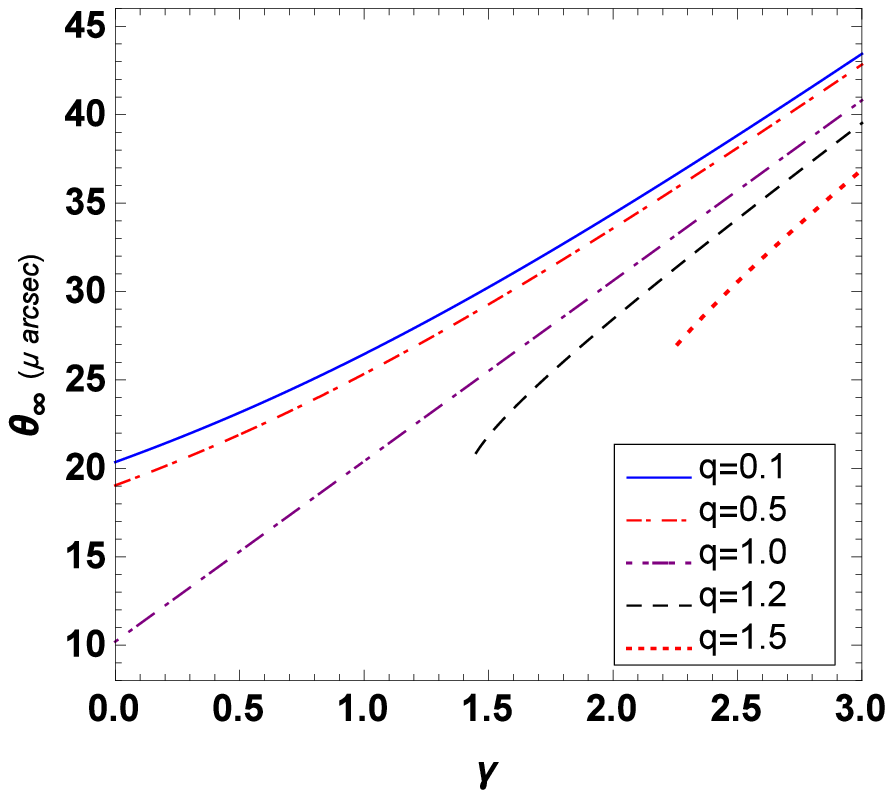}\\
\includegraphics[width=6cm]{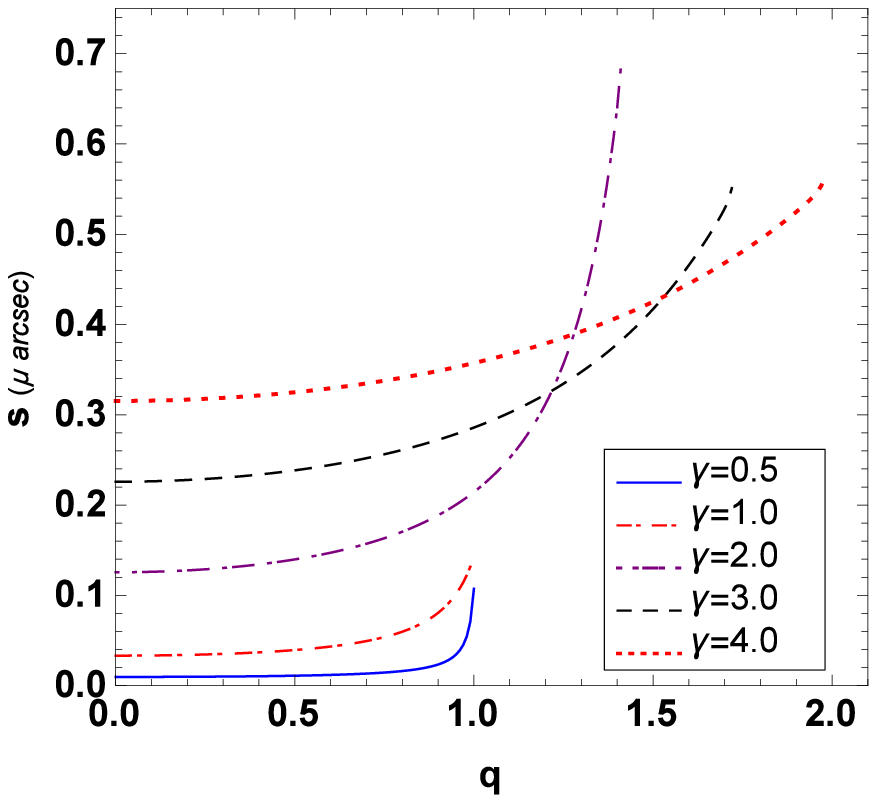}\includegraphics[width=6cm]{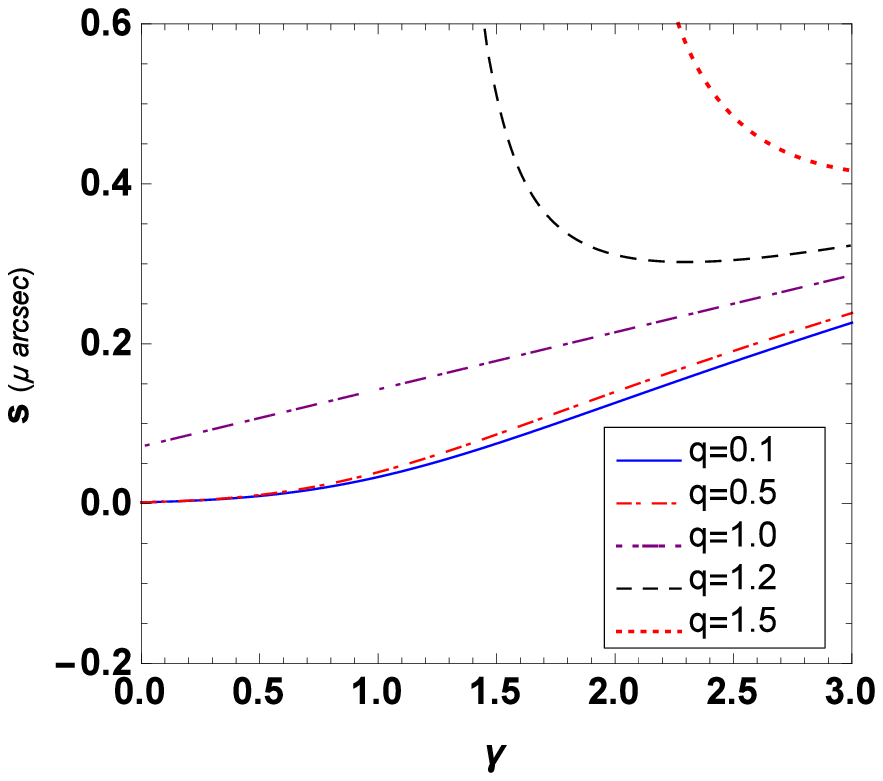}\\
\includegraphics[width=6cm]{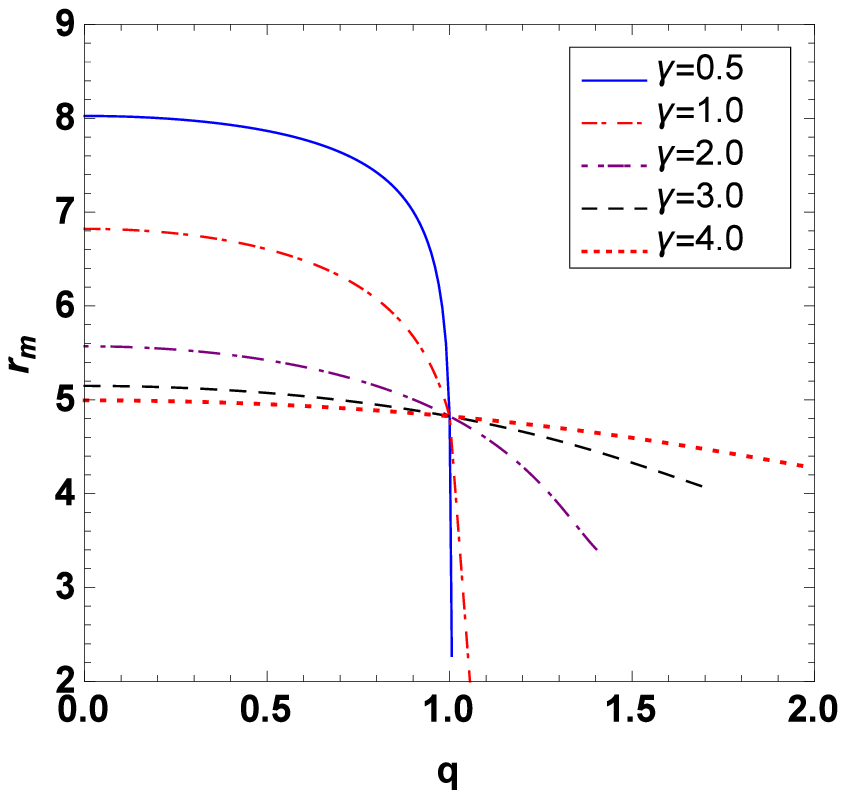}\includegraphics[width=6cm]{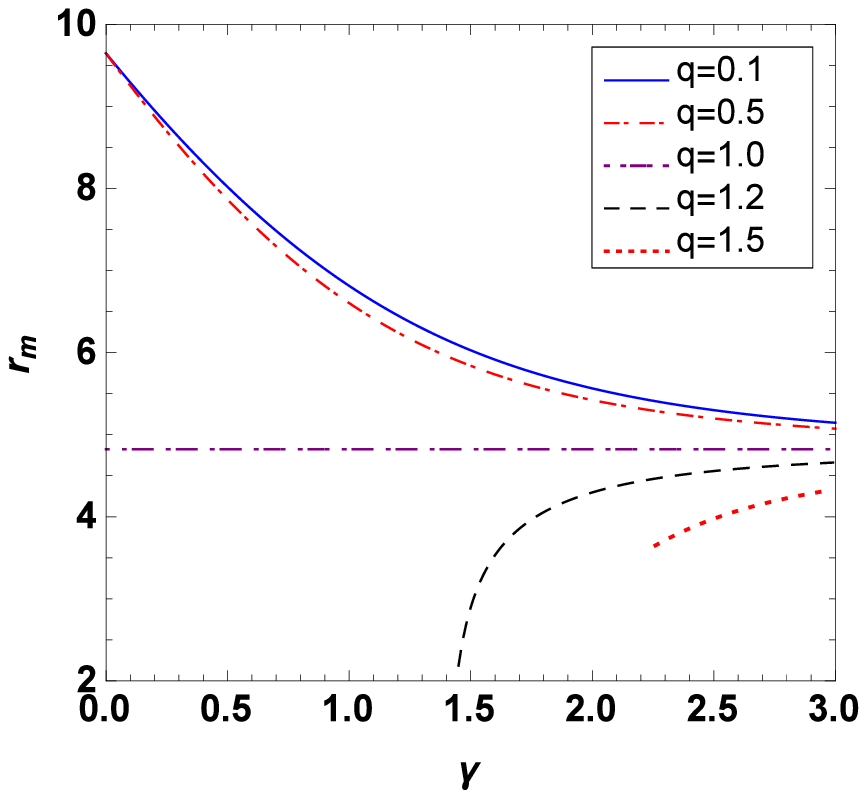}
\caption{Gravitational lensing by the center black hole in Milk Way Galaxy. Variation of the values of the angular position $\theta_{\infty}$, the angular separation $s$, and the relative magnitudes $r_m$ with the parameters $q$ and $\gamma$ in the spacetime with torsion (\ref{sol1}).}
\end{center}
\end{figure}
As the light source and observer are far enough from the central compact object, the corresponding lens equation can be expressed as  \cite{VB2}
\begin{eqnarray}
\bar{\gamma}=\frac{D_{OL}+D_{LS}}{D_{LS}}\theta-\alpha(\theta) \; mod
\;2\pi
\end{eqnarray}
where $D_{LS}$ is the lens-source distance and $D_{OL}$ is the
observer-lens distance. $\bar{\gamma}$ is the angle between the optical axis and the direction of the source. $\theta$ is the angular
separation between the lens and the image, which has a form $\theta=u/D_{OL}$. As in
Ref.\cite{VB2}, we here focus only on the simplest situation in which the source, lens and observer are highly aligned so that the angular separation between the lens and the $n$-th relativistic image can be simplified as
\begin{eqnarray}
\theta_n\simeq\theta^0_n\bigg(1-\frac{u_{ps}e_n(D_{OL}+D_{LS})
}{\bar{a}D_{OL}D_{LS}}\bigg),\label{nth0}
\end{eqnarray}
with
\begin{eqnarray}
\theta^0_n=\frac{u_{ps}}{D_{OL}}(1+e_n),\;\;\;\;\;\;
e_{n}=e^{\frac{\bar{b}+|\bar{\gamma}|-2\pi
n}{\bar{a}}},\label{nth1}
\end{eqnarray}
where $\theta^0_n$ is the image positions corresponding to
$\alpha=2n\pi$, and $n$ is an integer. As $n\rightarrow\infty$, from Eqs.(\ref{nth0}) and (\ref{nth1}), it is easy to find that
$e_n$ tends to zero, which means that the relationship between the asymptotic position of a set of images $\theta_{\infty}$ and the
minimum impact parameter $u_{ps}$ owns a simpler form
\begin{eqnarray}
u_{ps}=D_{OL}\theta_{\infty}.\label{uhs1}
\end{eqnarray}
In order to estimate the coefficients $\bar{a}$ and $\bar{b}$ as in Refs.\cite{VB1,VB2}, we can consider a perfect situation in which only the outermost image $\theta_1$ is separated as a single image and all the remaining ones are packed together at $\theta_{\infty}$. And then, the angular separation $s$ and the relative
magnitudes $r_m$ between the first image and other ones
can be simplified further as
\cite{VB1,VB2,Gyulchev1}
\begin{eqnarray}
s&=&\theta_1-\theta_{\infty}=\theta_{\infty}e^{\frac{\bar{b}-2\pi}{\bar{a}}},\nonumber\\
r_m&=&2.5\log{\frac{\mu_1}{\sum^{\infty}_{n=2}\mu_n}}
=\frac{5\pi}{\bar{a}}\log{e},\label{ss1}
\end{eqnarray}
Therefore, one can obtain the coefficients $\bar{a}$, $\bar{b}$ in strong deflection limit and the minimum impact parameter $u_{ps}$ through measuring $s$, $\theta_{\infty}$ and $r_m$ from observation experiments and extract further information about the central compact object. For Milk Way Galaxy, the mass of the central object of is estimated recently
to be $4.4\times 10^6M_{\odot}$ \cite{Genzel1} and its distance is
around $8.5kpc$, which leads to the ratio of the mass to the distance
$M/D_{OL} \approx2.4734\times10^{-11}$.  With these data,  we present the numerical value for the angular position of the relativistic images $\theta_{\infty}$, the angular separation $s$ and the relative magnitudes $r_m$ in Fig.(7). It is shown that the angular position of the relativistic images $\theta_{\infty}$ decreases with the parameter $q$ and increases with the parameter $\gamma$. The angular separation $s$ between $\theta_{1}$ and $\theta_{\infty}$ increases with the tidal-like charge $q$. With increase of $\gamma$, $s$ increases for the smaller $q$ case and decreases for the larger $q$ case. The change of $r_{m}$ with the parameters $q$ and $\gamma$ is converse to that of the angular separation $s$.

Finally, we make a brief comparison among the main features of strong gravitational lensing by compact objects in different gravitational theories, which is shown in Table (I).  It tells us that the event horizon and photon sphere are two important surfaces in the propagation of light rays.
From Table (I), it is obvious that in the cases with photon sphere the deflection angle $\alpha$ for the light ray near the photon sphere diverges logarithmically, which can be regarded as a common feature in the strong gravitational lensing by compact objects in various theories of gravity. In the cases of strong naked singularity where there is neither photon sphere nor horizon, the deflection angle $\alpha$ for the light ray near singularity tends to a finite value $\alpha_s$. The sign and value of $\alpha_s$ depend on the spacetime parameters.
\begin{table}
\begin{center}
\begin{tabular}{|c|c|c|c|c|c|c|}
\hline \hline $$ & \multicolumn{2}{c|}{Existence}&\multicolumn{3}{c|}{Deflection angle $\alpha$ close to} &	\\ \cline{2-6}
 Spacetimes& \footnotesize{ Event} & \footnotesize{Photon}&\footnotesize{ Event} & \footnotesize{Photon}&\footnotesize{}&Literature	\\
 & \footnotesize{horizon} & \footnotesize{sphere}& \footnotesize{horizon} & \footnotesize{sphere}&\footnotesize{Singularity} &	\\
\hline
\footnotesize{Schwarzschild black hole}& yes & yes &-& $\infty$&-& \cite{VB1,KS2}\\
\hline
\footnotesize{Reissner-N\"{o}rdstr\"{o}m black hole}& yes & yes &-& $\infty$&-& \cite{VB1,Eirc1}\\
\hline
 \footnotesize{Kerr black hole} &yes & yes &-& $\infty$&-&\cite{VB2,VB202} \\
\hline
\multirow{2}{*}{\footnotesize{Janis-Newman-Winicour naked singularities}}  &no &yes &-& $\infty$&-&  \multirow{2}{*}{\cite{KS4,Gyulchev1}}\\ \cline{2-6}
 &no& no &-& -& \footnotesize{finite value} & \\
 \hline
\footnotesize{Dilaton black holes}& yes & yes &-& $\infty$&-& \cite{Gyulchev,Bhad1,TSa1}
\\
\hline
\footnotesize{Ho\v{r}ava-Lifshitz black holes}& yes & yes &-& $\infty$&-& \cite{Song1,Song1add}
\\
\hline
\footnotesize{Squashed Kaluza-Klein black holes}& yes & yes &-& $\infty$&-& \cite{Song101,gr4,gr401}
\\
 \hline
\footnotesize{Braneworld black holes}& yes & yes &-& $\infty$&-& \cite{whisk,AnAv,bran1}
 \\
 \hline
\footnotesize{Eddington-inspired Born-Infeld black holes}& yes & yes &-& $\infty$&-& \cite{agl1,agl1add}
\\
\hline
\footnotesize{Einstein-Born-Infeld black holes}& yes & yes &-& $\infty$&-& \cite{agl1add0}
 \\
  \hline
\footnotesize{Black Holes in Loop Quantum Gravity}& yes & yes &-& $\infty$&-& \cite{agl101}
 \\
 \hline
\footnotesize{Black hole pierced by a cosmic string} & yes & yes &-& $\infty$&-&\cite{gr201}
 \\
 \hline
\footnotesize{Kerr-Taub-NUT spacetime} & yes & yes &-& $\infty$&-&\cite{gr2}
 \\
 \hline
\footnotesize{Gauss-Bonnet black hole} & yes & yes &-& $\infty$&-&\cite{gr51}
 \\
 \hline
\multirow{2}{*}{\footnotesize{Johannsen-Psaltis rotating non-Kerr spacetime}} & yes & yes &-& $\infty$&-&\multirow{2}{*}{\cite{Song102}}\\ \cline{2-6}
 &no& no &-& -& \footnotesize{finite value} &
 \\
 \hline
\multirow{3}{*}{\footnotesize{Konoplya-Zhidenko rotating non-Kerr spacetime}} & yes & yes &-& $\infty$&-&\multirow{3}{*}{\cite{Song2}}\\ \cline{2-6}
 &no& yes &-& $\infty$& - &
 \\ \cline{2-6}
 &no& no &-& -& \footnotesize{finite value} &
 \\
 \hline
\footnotesize{Phantom black holes} & yes & yes &-& $\infty$&-&\cite{gr1,gr1add0,gr1add}
 \\
 \hline
\footnotesize{Noncommutative black hole} & yes & yes &-& $\infty$&-&\cite{gr1noncom}
 \\
 \hline
\footnotesize{Einstein-Skyrme black hole} & yes & yes &-& $\infty$&-&\cite{glcom31}
 \\
 \hline
\footnotesize{Horndeski black hole}& yes & yes &-& $\infty$&-&\cite{glcom32,glcom33}\\
\hline
\footnotesize{Black hole in massive gravity}& yes & yes &-& $\infty$&-&\cite{mg1}\\
\hline
\footnotesize{Black hole in Chern-Simons modified gravity}& yes & yes &-& $\infty$&-&\cite{CSmg}\\
\hline
\footnotesize{Black hole in F(R) gravity}& yes & yes &-& $\infty$&-&\cite{fR1,fR2}\\
\hline
\footnotesize{Wormholes}& no & yes &-& $\infty$&-&\cite{worm1,worm2,worm3}
\\
\hline
\footnotesize{High dimensional black holes}& no & yes &-& $\infty$&-&\cite{Exdim1,Exdim2}\\\hline
\footnotesize{Charged black holes in scalar-tensor gravity}& no & yes &-& $\infty$&-&\cite{Scatensor}\\
\hline
\multirow{3}{*}{\footnotesize{Bardeen spacetime}} & yes & yes &-& $\infty$&-&\multirow{3}{*}{\cite{Bardeen}}\\ \cline{2-6}
 &no& yes &-& $\infty$& - &
 \\ \cline{2-6}
 &no& no &-& -& \footnotesize{finite value} &
 \\
 \hline\multirow{4}{*}{\footnotesize{Spacetime with torsion \cite{sbh1} in ECKS theory}}& yes & yes &-& $\infty$&-&\multirow{4}{*}{\cite{zchen}}\\ \cline{2-6}
 &no&yes &-& $\infty$& - & \\\cline{2-6}
 &yes& no &$\infty$& -& - & \\\cline{2-6}
 &no& no &-& -& \footnotesize{ finite value} & \\
  \hline\multirow{4}{*}{\footnotesize{ Spacetime with torsion (\ref{sol1}) in ECKS theory}}& yes & yes &-& $\infty$&-&\multirow{4}{*}{In our current work}\\
  \cline{2-6}
 &no&yes &-& $\infty$& - & \\\cline{2-6}
 &yes& no &\footnotesize{finite value} & -& - & \\\cline{2-6}
 &no& no &-& -& \footnotesize{finite value} & \\
\hline\hline
\end{tabular}
\end{center}
\label{tab1} \caption{Comparison among the main features of gravitational lensing by compact objects in different gravitational theories.}
\end{table}
From Table.(I), we also note that only in the spacetimes with a torsion in generalized ECKS theory considered here or in Ref.\cite{zchen},  there is a special parameter region in which there exists only horizon but no photon sphere. In this special parameter region, the deflection angle $\alpha$ is also logarithmically divergent as the light ray approaches the event horizon in the spacetime \cite{sbh1}. However, for the spacetime (\ref{sol1}), we find that the deflection angle $\alpha$ tends to a certain finite value as the light ray approaches the event horizon, which could be understand by a fact that the photon is captured directly by black hole before it make infinite complete loops around the central object. For the black holes in general relativity (i.e., the coupling coefficient $a_1=0$ ), such as, Schwarzschild and Kerr black holes, one can find that the photon sphere lies always outside of the event horizon and then for the light ray from infinity the deflection angle $\alpha$  diverges logarithmically as the closest approach distance $r_0$ tends to the photon sphere, which means that photons are captured by black hole once they reach the photon sphere. However, in the generalized ECKS theory, there are a special kind of black hole without photon sphere  (\ref{sol1}) or in Ref.\cite{sbh1}, so that  photons can be captured by black hole only when they reach the event horizon, which differs from that in the black holes in general relativity and other theories of gravity. Moreover, comparing with in Reissner-N\"{o}rdstr\"{o}m black hole case, as $\gamma>1$,  due to the larger coefficient $\bar{a}$, the deflection angle for light ray with the same $r_0$ in the strong field limit is larger in the spacetime of the black hole with photon sphere (\ref{sol1}). Similarly, the observables $\theta_{\infty}$ and  $s$ have larger values. However, as $\gamma<1$, the deflection angle, the observables $\theta_{\infty}$ and  $s$ are less than that in Reissner-N\"{o}rdstr\"{o}m black hole case. On the contrary, comparing with in Reissner-N\"{o}rdstr\"{o}m black hole spacetime, the relative magnitudes $r_m$ is smaller for the cases with $\gamma>1$ and is larger for the cases with $\gamma<1$.
These deviation from the cases in general relativity could help us in understanding the gravitational lensing caused by torsion in the generalized ECKS theory of gravity.

\section{summary}

In this paper we firstly present a new black hole solution in the generalized ECKS gravity with three independent parameters $m$, $q$ and $\gamma$ and then investigate the propagation of photon in this background. We find that these spacetime parameters affect sharply photon sphere, deflection angle of light ray and strong gravitational lensing. The photon sphere exists only in the regime $q<q_c$ and the value of $q_c$ depends on the parameter $\gamma$. In the regime where photon sphere exists, the radius of photon sphere increases with
the parameter $\gamma$, but decreases with $q$. Moreover, the condition of existence of horizons is not inconsistent with that of photon sphere, which yields that the whole region in the panel ($\gamma, q$) can be split into four regions by the
boundaries of the existence of horizon and of the photon sphere.
In the cases with photon sphere,
the deflection angle of the light ray near the photon sphere diverges logarithmically, which is similar to those in the usual spacetime of a black hole or a weak naked singularity in the strong-field limit. In the case without photon sphere and horizon,
the deflection angle of the light ray closing very to the singularity approaches a finite value $-\pi$, which does not depend on spacetime parameters $\gamma$ and $q$. It should be a common feature of the deflection angle of light ray near the static strong naked singularity.
Furthermore, we also find that there exists a special case in which there is horizon but no photon sphere for the spacetime (\ref{sol1}) as in the spacetime with a torsion \cite{zchen}.
However, we find that the deflection angle of the light ray near the event horizon tends to a finite value in this case, which differs from those in the black hole with a torsion considered in Ref.\cite{zchen} where the deflection angle of the light finally becomes diverges logarithmically. It could be attribute to that the photon is captured directly by black hole before it make infinite complete loops around the central object in this case.
Finally, we studied the strong gravitational lensing by a compact object (\ref{sol1}) with the photon sphere and then probed how the parameters $\gamma$ and $q$
affect the coefficients in the strong field limit.

\section{\bf Acknowledgments}
We would like to thank the referee for useful comments.
This work was partially supported by the National Natural Science Foundation of China under
Grant No. 11875026, the Scientific Research
Fund of Hunan Provincial Education Department Grant
No. 17A124. J. Jing's work was partially supported by
the National Natural Science Foundation of China under
Grant No. 11475061, 11875025.

\end{document}